\documentclass[journal]{IEEEtran}

\usepackage{graphicx}
\usepackage{float}
\usepackage{mathrsfs}
\usepackage{amsfonts}
\usepackage{cite}
\usepackage{color}
\usepackage{subfigure}
\usepackage{booktabs}
\usepackage{multirow}
\usepackage{algorithm}
\usepackage{algpseudocode}
\usepackage{graphics}

\ifCLASSINFOpdf
\else
\fi
\begin{document}

\title{Optimal Rate-Diverse Wireless Network Coding }
\author{Taotao Wang, \emph{Member, IEEE}, Soung Chang Liew, \emph{Fellow, IEEE}, and Long Shi, \emph{Member, IEEE} 

\thanks{Taotao Wang and Soung Chang Liew are with the Institute of Network Coding, The Chinese University of Hong Kong (email: \{ttwang, soung,\}@ie.cuhk.edu.hk). Long Shi is with the College of Electronic and Information Engineering, Nanjing University of Aeronautics and Astronautics, Nanjing, China (email: slong1007@gmail.com). }
}
\markboth{Optimal Rate-Diverse Wireless Network Coding}%
{Shell \MakeLowercase{\textit{et al.}}: Bare Demo of IEEEtran.cls
for Journals}
\maketitle

\begin{abstract}
This paper proposes an encoding/decoding framework for achieving the optimal channel capacities of the two-user broadcast channel where each user (receiver) has the message targeted for the other user (receiver) as side information. Since the link qualities of the channels from the base station to the two users are different, their respective single-user non-broadcast channel capacities are also different.  A goal is to simultaneously achieve/approach the single-user non-broadcast channel capacities of the two users with a single broadcast transmission by applying network coding. This is referred to as the \emph{rate-diverse wireless network coding}  problem. For this problem, this paper presents a capacity-achieving framework based on linear- structured nested lattice codes. The significance of the proposed framework, besides its theoretical optimality, is that it suggests a general design principle for linear rate-diverse wireless network coding going beyond the use of lattice codes. We refer to this design principle as the \emph{principle of virtual single-user channels}. Guided by this design principle, we propose two implementations of our encoding/decoding framework using practical linear codes amenable to decoding with affordable complexities: the first implementation is based on Low Density Lattice Codes (LDLC) and the second implementation is based on Bit-interleaved Coded Modulation (BICM). These two implementations demonstrate the validity and performance advantage of our framework.
\end{abstract}


\IEEEpeerreviewmaketitle

\section{Introduction}


\IEEEPARstart{T}{his} paper investigates wireless broadcast networks with side information at users, where a base station wants to deliver two different messages to two users A and B, and each user already has the message targeted for the other user as side information. Two examples in which this scenario can arise are as follows: 

Scenario 1: Two-way Relay Networks  ---  In this scenario \cite{rankov2007spectral, zhang2006hot}, the message for B originates from A; and the message for A originates from B.  These messages have been transmitted by users A and B to the base station via a prior uplink phase. The base station serves as a relay to deliver the information to the respective destinations, B and A,  in the downlink phase. The current paper is related to the downlink phase in this setting. 

Scenario 2: Retransmission in Automatic Repeat reQuest (ARQ) --- In this scenario, the information originates elsewhere (e.g., the Internet). Originally, neither A nor B had information on the messages for the other nodes. However, in prior downlink transmissions, A overheard the message for B, and B overheard the message for A, but neither A or B succeeded in decoding their own desired messages. Retransmissions will be needed. In a future downlink retransmission, both A and B have side information on the messages for the other user. 

Besides the above two scenarios, there are many other scenarios in which downlink broadcast with side information arise \cite{asadi2014index}.  Network coding \cite{ahlswede2000network, li2003linear} can naturally be used to improve the transmission efficiency for such wireless broadcast networks. With network coding, the base station transforms the two messages into one network-coded message and sends the network-coded message to the two users. Each user then decodes its desired message by subtracting the side information from the received network-coded message. 

In this paper, we assume the channels from the base station to the two users are power-constrained additive white Gaussian noise (AWGN) channels. The capacities of the two channels can be different due to the different channel qualities from the base station to two users. We refer to the coding problem for such channels as the rate-diverse wireless network coding problem, because the achievable rates for the two channels are different; yet our goal is to be able achieve the respective rates with a single transmission of the network-coded message. Specifically, of concern are the following two questions:
\begin{itemize}
	\item[i)] What is the capacity region of rate-diverse wireless network coding?
	\item[ii)] How to achieve (or approach) the optimal operating point of the capacity region using practical and implementable coding schemes?
\end{itemize}

There has been prior works addressing question i). The capacity region of the channel under investigation has been identified in \cite{wu2007broadcasting, kramer2007capacity, oechtering2008broadcast} using the argument of random coding; it was proved that the optimal point of the capacity region is the pair of the two point-to-point single-user channel capacities. However, based on the theoretical arguments in \cite{wu2007broadcasting, kramer2007capacity, oechtering2008broadcast} alone, the answer to question ii) is not obvious.  This paper is oriented toward question ii). We find that simple coding schemes can allow us to achieve/approach the point-to-point channel capacities, using a principle put forth by us, referred to as the principle of virtual single-user channels.

To achieve the optimal point of the capacity region for rate-diverse wireless network coding in power-constrained AWGN channels, we propose an encoding/decoding framework based on linear nested lattice codes \cite{erez2004achieving}. The merit of our framework is twofold. First, it shows that we can achieve the optimal point of the capacity region using linear structured codes. Second, it yields a general design principle for rate-diverse wireless network coding based on linear codes. We refer to this design principle as the principle of virtual single-user channels. This principle indicates that there is no need to perform joint two-user encoding to achieve capacities. Separate channel coding before network coding at the transmitter, and single-user decoding after network decoding at the receiver, are sufficient to achieve capacities.  

Importantly, this principle is not just limited to the use of nested lattice codes (although the insight originally came from the use of nested lattice codes). For example, we can implement our encoding/decoding framework using practical linear codes amenable to decoding with affordable complexities. In other words, although the framework originates from the use of theoretical lattice codes with infinite dimensions, the design principle can be easily applied in practice using other linear codes.  

To illustrate, this paper shows how the optimal encoding/decoding framework can be applied using low density lattice codes (LDLC) \cite{LDLC2008}.  Although the nested lattice code is optimal in the sense of capacity-achieving, its lattice quantization decoding is not computationally feasible. LDLC is a type of lattice codes amenable to low-complexity belief propagation (BP) decoding \cite{yedidia2003understanding, kschischang2001factor}. Using LDLC as components, our encoding/decoding framework for rate-diverse wireless network coding can be easily realized in practical systems.

This paper further illustrates the application of the framework using bit-interleaved coded modulation (BICM) \cite{caire1998bit}, which has even lower implementation complexity than LDLC.  BICM is already used in many practical point-to-point communication systems to approach capacity in the high signal-to-noise ratio (SNR) regime. BICM combines simple binary channel codes with high-order modulations to obtain different data rates in accordance with the SNR. This paper shows that applying the principle of virtual single-user channels allows both users to achieve their respective single-user point-to-point decoding performance in BICM-based rate-diverse wireless network coding systems. This result is contrast to prior works \cite{chen2010novel, yun2010rate}, where more complicated joint modulation schemes are adopted. These joint modulation schemes are designed for specific modulations and are difficult to generalize. More importantly, as will be seen in this paper, they do not allow optimal single-user decoding performance to be simultaneously achieved for both users.

In summary, this paper has the following three contributions:

\begin{itemize}
	\item We put forth a framework based on structured nested lattice codes to achieve the optimal capacities in rate-diverse wireless network coding systems. This framework provides insight leading to a general design principle of using linear codes to achieve optimal rate-diverse network coding performance, referred to as the principle of virtual single-user channels. 
	
	\item We put the design principle into practice assuming the use of LDLC that are amenable to practical encoding/decoding using BP algorithms. We demonstrate the good performance of the resulting system.
	
	\item We further apply the design principle assuming the use of BICM. The proposed BICM scheme can achieve the single-user performance bounds for the two users simultaneously. 
	
\end{itemize}

\subsection{Related Works}

Ref. \cite{schnurr2007coding} considered the use of linear codes to achieve the optimal point of the capacity region in finite-alphabet channels (channels with discrete outputs). By contrast, our paper here focuses on power-constrained additive white Gaussian noise (AWGN) channels, which corresponds more closely to the actual physical channel at the lower layer. The coding scheme in \cite{xiao2006nested} is a solution for rate-diverse wireless network coding in power-constrained AWGN channels; since it only employs binary linear codes and BPSK modulation, it cannot achieve capacity in all SNR regimes. 

The constellations proposed in \cite{natarajan2014lattice} for noisy index coding can be also applied to rate-diverse wireless network coding. However, it aims to achieve the so-called side information gain but not to achieve the optimal capacity point. 

To boost the performance of rate-diverse wireless network coding systems, \cite{chen2010novel} and \cite{yun2010rate} proposed two-user joint modulation schemes for non-channel-coded systems. However, these non-systematic designs for joint modulation schemes cannot achieve the optimal single-user performance for the two users simultaneously.  

A recent independent work \cite{natarajan2015lattice,natarajan2015latticeful} also employed the idea of nested lattice codes to achieve the capacity of the Gaussian broadcast channel. In \cite{natarajan2015lattice,natarajan2015latticeful}, a more general model with more than two users was considered, and a theoretical optimal solution was given for the general model. However, for the case of two users, our work and \cite{natarajan2015lattice,natarajan2015latticeful} differ in the following ways. First, our encoding scheme employs two different coding lattices to independently encode the two messages, where each coding lattice is exactly the same as that used in traditional single-user point-to-point communication; the encoding scheme of \cite{natarajan2015lattice, natarajan2015latticeful}, on the other hand, jointly encodes the messages into a single coding lattice. Second, our decoding scheme, after the network-coding decoding operation, is the lattice quantization operation with respect to the original single-user coding lattice; the decoding scheme of\cite{natarajan2015lattice, natarajan2015latticeful}, on the other hand, is a lattice quantization with respect to a new coding lattice representing a sub-code of the original lattice code. Our scheme suggests a simple design principle for the two-user case: the principle of virtual single-user channels. This principle dictates that we should aim for a design in which the whole decoding problem can be reduced to two uncoupled single-user decoding problems after the side information at the receivers are taken into account. Importantly, for a design adhering to this principle, we can simply use single-user channel codes, including non-lattice codes (e.g., LDPC codes) that are capacity approaching yet amenable to simple decoding.  

Overall, despite the past related works, how to achieve the two point-to-point single-user channel capacities simultaneously using structured codes is not obvious for rate-diverse wireless network coding in power-constrained AWGN channels. It is the intention of this paper to fill this gap.  

The rest of this paper is organized as follows. Section II describes our system model. Using the argument of nestled lattice codes, Section III presents our optimal encoding/decoding framework for rate-diverse wireless network coding. Section IV describes the implementation of our encoding/decoding framework using LDLC. Section V proposes the BICM scheme for rate-diverse wireless network coding. Section VI provides simulation results. Section VII concludes this paper.

\section{System Model}

\begin{figure}[!t]
	\centering
	\includegraphics[width=3in]{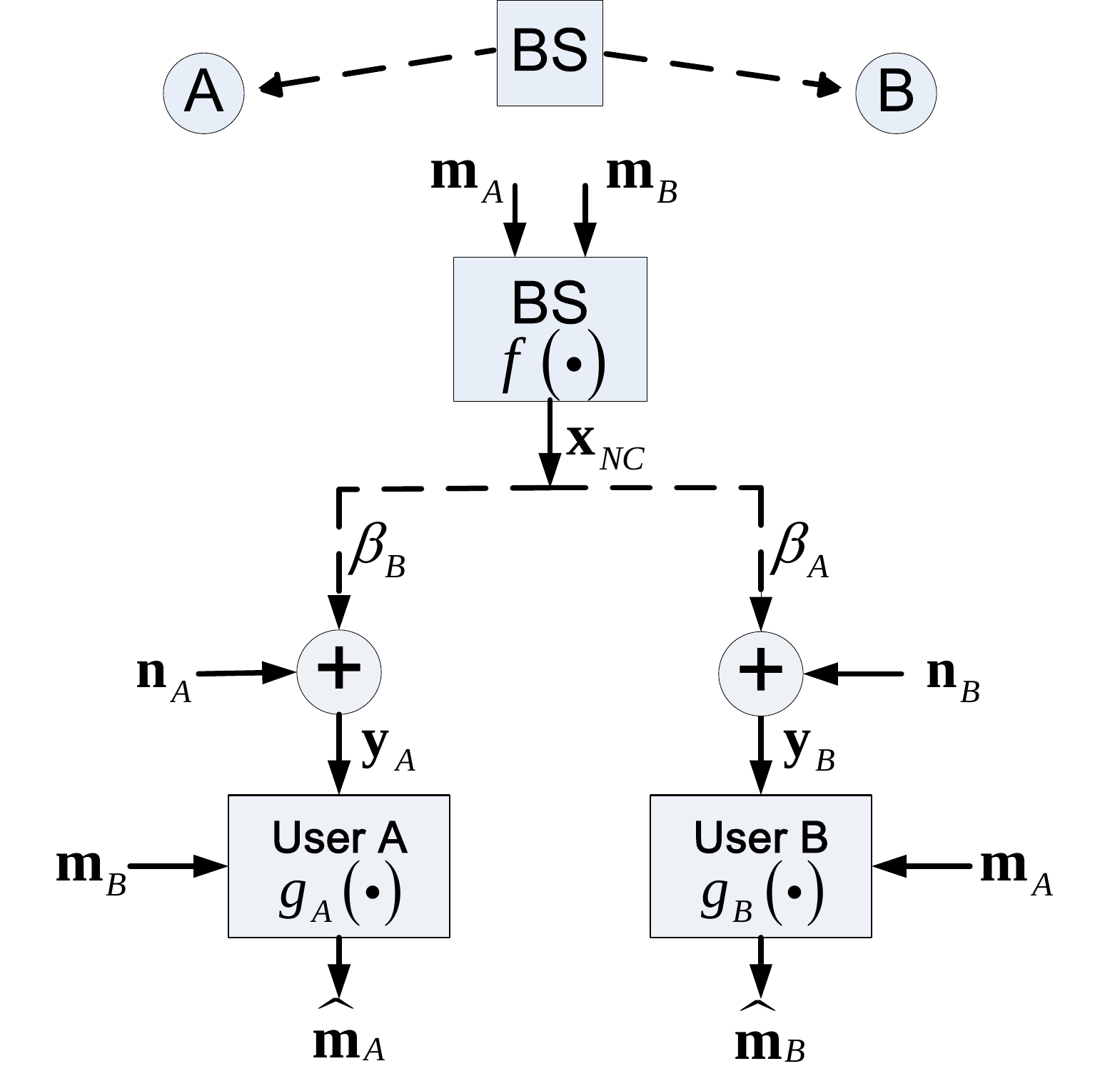}
	\caption{The system model of the broadcast channel with side-information at the users.} \label{block_system_model}
	\vskip -0.2in
\end{figure}

We consider a network-coding assisted wireless broadcast problem. The system model is shown in Fig.~\ref{block_system_model}, where we have a base station (BS) and two users (A and B). The BS wants to transmit different messages to users A and B. The message targeted for user A (B) is denoted by a vector of binary information bits ${{\mathbf{m}}_A} \in {\left\{ {0,1} \right\}^{{L_A}}}$ (${{\mathbf{m}}_B} \in {\left\{ {0,1} \right\}^{{L_B}}}$), where $L_A$ and $L_B$ are the lengths of the vectors. User A (B) has side information ${{\mathbf{m}}_B}$ (${{\mathbf{m}}_A}$), the message targeted for user B (A). This is a common scenario in the broadcast phase of the two-way relay channel based on physical-layer network coding \cite{zhang2006hot}; and this is also a special case of index coding systems \cite{asadi2014index}.

The BS can employ network coding to minimize the required transmission time \cite{ahlswede2000network, li2003linear}. Besides network coding, the BS also needs to perform channel coding and modulation to generate the channel symbols that will be transmitted over the wireless channel. For simplicity, this paper considers a real-valued signal model. The extension to a complex-valued signal model is straightforward. We denote the vector of channel symbols transmitted by the BS by ${{\mathbf{x}}_{NC}} = f\left( {{{\mathbf{m}}_A},{{\mathbf{m}}_B}} \right)$, where ${{\mathbf{x}}_{NC}} \in {\mathbb{R}^N}$ consists of $N$ channel symbols, and the function $f\left(  \cdot 
\right)$ incorporates the combined operation of channel coding, modulation and network coding. Thus, the data rate for user A (B) is ${R_A} = {{{L_A}} \mathord{\left/
		{\vphantom {{{L_A}} N}} \right.
		\kern-\nulldelimiterspace} N}$ (${R_B} = {{{L_B}} \mathord{\left/
		{\vphantom {{{L_B}} N}} \right.
		\kern-\nulldelimiterspace} N}$) bits per channel use. We impose an average power constraint ${P_X}$ on the channel symbols, i.e., ${{E{{\left\| {{{\mathbf{x}}_{NC}}} \right\|}^2}} \mathord{\left/
		{\vphantom {{E{{\left\| {{{\mathbf{x}}_{NC}}} \right\|}^2}} N}} \right.
		\kern-\nulldelimiterspace} N} \le {P_X}$.

We model the wireless channel between the BS and user $u \in \left\{ {A,B} \right\}$ as an additive white Gaussian noise (AWGN) channel with the path-loss effect:
\begin{equation}
	{{\bf{y}}_u} = {\beta _u}{{\bf{x}}_{NC}} + {{\bf{n}}_u}
\end{equation}
where ${{\bf{n}}_u} \in \mathbb{R}^N$ is a vector of i.i.d. mean-zero, variance-$\sigma _n^2$ Gaussian white noise components, and $0 < {\beta _u} \in \mathbb{R} $
is the channel gain that models the path-loss effect between the BS and user $u$. The two channel gains ${\beta_A}$ and ${\beta_B}$ are likely different due to the different distances of the users from the BS.  

After receiving ${{\bf{y}}_A}$ (${{\bf{y}}_B}$), user A (B) estimates its target message ${{\mathbf{m}}_A}$ (${{\mathbf{m}}_B}$) using ${{\bf{y}}_A}$ (${{\bf{y}}_B}$) and its side information ${{\mathbf{m}}_B}$ (${{\mathbf{m}}_A}$). Specifically, we express the estimated target messages as ${\widehat{\bf{m}}_A} = {g_A}\left( {{{\bf{y}}_A},{{\bf{m}}_B}} \right)$, ${\widehat{\bf{m}}_B} = {g_B}\left( {{{\bf{y}}_B},{{\bf{m}}_A}} \right)$, 
where ${g_A}\left( \cdot \right)$ and ${g_B}\left( \cdot \right)$ denote the combined inverse operation of channel coding, modulation and network coding at users A and B, respectively. Henceforth, for brevity, we will simply call $f\left( \cdot \right)$ the encoding scheme, and  ${g_A}\left( \cdot \right)$ and ${g_B}\left( \cdot \right)$ the decoding schemes. Given the average power constraint over the channel symbols of BS, we ask the following two questions:
\begin{itemize}
	\item[i)] \emph{What are the data-rate limits for the BS to reliably deliver messages to the two users?}
	\item[ii)] \emph{What encoding/decoding schemes can be used to achieve these limits?}
\end{itemize}

\begin{figure}[!t]
	\centering
	\includegraphics[width=2in]{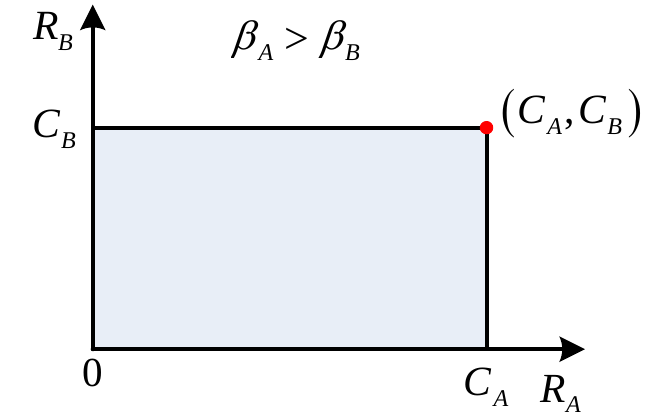}
	\caption{The capacity region of the broadcast channel with side-information at the two users.} \label{cap_re}
	\vskip -0.1in
\end{figure}

We first identify the data-rate limits for the channel considered here. If we just focus on the point-to-point single-user channel between the BS and one particular user $u$, the Shannon channel capacity, ${C_u} \buildrel \Delta \over = \left( {{1 \mathord{\left/
			{\vphantom {1 2}} \right.
			\kern-\nulldelimiterspace} 2}} \right){\log _2}\left( {1 + SN{R_u}} \right)$, is the upper limit of the data rate for which reliable communication is possible as $N \to \infty $, where $SN{R_u} = {{{P_X}\beta _u^2} \mathord{\left/
		{\vphantom {{{P_X}\beta _u^2} {\sigma _n^2}}} \right.
		\kern-\nulldelimiterspace} {\sigma _n^2}}$ is the SNR at the receiver of user $u$. Considering the broadcast channel with side-information at the two users, references  \cite{wu2007broadcasting, kramer2007capacity, oechtering2008broadcast} proved that as long as the date-rate pair $\left( {{R_A},{R_B}} \right)$ is within the capacity region given by $\left\{ {\left( {{R_A},{R_B}} \right):{R_A} < {C_A},{R_B} < {C_B}} \right\}$, the users can decode their target messages with arbitrarily small error probabilities. The capacity region is shown in Fig.~\ref{cap_re}, where we assume ${\beta _A} > {\beta _B}$. Obviously, the capacity pair $\left( {{C_A},{C_B}} \right)$ is the optimal data-rate pair that simultaneously maximizes the data rates for both users.  

This paper focuses on the second question: the encoding and decoding schemes to achieve the optimal data-rate pair (the capacity pair). Let us first consider the simple special case where the two channel gains are equal: ${\beta _A} = {\beta _B}$. Now, the point-to-point channel capacities of the two channels are equal: ${C_A} = {C_B} = C$, and the capacity pair becomes $\left( {C,C} \right)$. For this rate-equal case, the encoding scheme for capacity achieving is rather straightforward. The first step is the linear network coding over the binary information: ${{\bf{m}}_{NC}} = {{\bf{m}}_A} \oplus {{\bf{m}}_B}$, where $\oplus$ denotes the bit-wise XOR operation.  Then, the network-coded message ${{\bf{m}}_{NC}}$ is fed into a single-user channel encoder and modulator. At the receiver side of a user, a single-user decoding scheme can be used to obtain ${\widehat{\bf{m}}_{NC}}$, and the estimated target message is given by the bit-wise XOR of ${\widehat{\bf{m}}_{NC}}$ and the side-information. As long as the data rate of the used encoding scheme $R$ can achieve the point-to-point single-user channel capacity $C$, the above simple encoding scheme can also achieve the capacity pair $\left( {C,C} \right)$ for the rate-equal wireless network coding.

Of interest to our paper here is the general rate-diverse wireless network coding where ${\beta _A} \ne {\beta _B}$ and ${C_A} \ne {C_B}$. How to achieve the capacity pair $\left( {{C_A},{C_B}} \right)$ for the rate-diverse case is not as obvious as the rate-equal case. In \cite{oechtering2008broadcast}, random coding is employed to derive the capacity region of the general probabilistic broadcast channel $p\left( {{{\bf{y}}_A},{{\bf{y}}_B}\left| {{{\bf{x}}_{NC}}} \right.} \right)$ with side information at the users. Then, \cite{schnurr2007coding} considered the use of linear codes to achieve the optimal capacity pair for finite-alphabet channels (channels with discrete outputs). By contrast, our paper here focuses on power-constraint AWGN channels. We investigate the encoding scheme and practical decoding scheme with affordable complexities to achieve (or to closely approach) the capacity pair. In particular, we put forth a design principle for rate-diverse wireless network coding. The design principle, referred to as \emph{the principle of virtual single-user channels}, aims to transform the rate-diverse broadcast channel to two single-user channels states through encoding/decoding designs, thereby achieving the capacity pair which are basically the single-user capacities for user  A and user B.

\section{Nested Lattice Codes Based Framework for Achieving Capacity Pair}

This section describes an encoding/decoding framework for achieving the capacity pair of rate-diverse wireless network coding. The framework is based on nested lattice codes that have linear structures. We first give a preliminary on lattices and nested lattice codes in Section III.A. Then, in Section III.B, we show how nested lattice codes and its decoding can be used in our encoding/decoding framework to achieve the capacity pair.

\subsection{Preliminary on Lattices and Nested Lattice Codes}

A real \emph{lattice} $\Lambda $ of dimension $K$ is a discrete subgroup of $\mathbb{R}^N$ ($K \le N$) closed under addition and reflection: if ${{\bf{\lambda }}_1},{{\bf{\lambda }}_2} \in \Lambda $, then ${{\bf{\lambda }}_1} + {{\bf{\lambda }}_2} \in \Lambda $; if ${\bf{\lambda }} \in \Lambda $ then $-{\bf{\lambda }} \in \Lambda $. The lattice points (vectors) of $\Lambda $ are generated by taking all integer linear combinations of $K$ independent basis vectors. The $K$ basis vectors can be written into an $N \times K$ generator matrix ${\bf{G}} \in {\mathbb{R}^{N \times K}}$. Therefore, a lattice $\Lambda $  is specified by its generator matrix  $\bf{G}$  and can be always written as $\Lambda \left( {\bf{G}} \right) \buildrel \Delta \over = \left\{ {{\bf{\lambda }} = {\bf{Gb}}:{\bf{b}} \in {\mathbb{Z}^K}} \right\}.$

Given a lattice $\Lambda $ and a lattice point  ${\bf{\lambda }} \in \Lambda $, the \emph{Voronoi region} of ${\bf{\lambda }} $ is defined to be the set of all vectors in $\mathbb{R}^N$ that are closest to the lattice point ${\bf{\lambda }} $: ${\cal V}\left( {\Lambda ,{\bf{\lambda }}} \right) \buildrel \Delta \over = \left\{ {{\bf{x}} \in {\mathbb{R}^N}:\left\| {{\bf{\lambda }} - {\bf{x}}} \right\| < \left\| {{\bf{\lambda '}} - {\bf{x}}} \right\|,\forall {\bf{\lambda '}} \in \Lambda ,{\bf{\lambda '}} \ne {\bf{\lambda }}} \right\}$. The Voronoi region of ${\bf{\lambda }} = {\bf{0}}$ is called the fundamental Voronoi region of the lattice and it is denoted by ${\cal V}$. Since every vector in $\mathbb{R}^N$  can be uniquely written as ${\bf{x}} = {\bf{\lambda }} + {\bf{r}}$, where ${\bf{\lambda }} \in \Lambda $ and ${\bf{r}} \in {\cal V}$, a \emph{lattice quantization}  operation is a function that maps a vector ${\bf{x}} \in {\mathbb{R}^N}$ to a lattice point of $\Lambda $ according to the minimum Euclidean distance rule: ${Q_\Lambda }\left( {\bf{x}} \right) \buildrel \Delta \over = \arg \mathop {\min }\nolimits_{{\bf{\lambda '}} \in \Lambda } \left\| {{\bf{x}} - {\bf{\lambda '}}} \right\| = {\bf{\lambda }};$   a \emph{lattice modulo} operation is to get the quantization error:  
${\bf{x}}\;{\rm{ mod }}\Lambda  \buildrel \Delta \over = {\bf{x}} - {Q_\Lambda }\left( {\bf{x}} \right) = {\bf{r}}$.

The volume of ${\cal V}$ is denoted by ${\rm{Vol}}\left( {\cal V} \right)$ and it can be shown that   ${\rm{Vol}}\left( {\cal V} \right) = \int_{\cal V} {d{\bf{x}}}  = \sqrt {\det \left( {{{\bf{G}}^T}{\bf{G}}} \right)} $.  The second-order moment of ${\cal V}$  is defined as ${\sigma ^2}\left( \cal V  \right) \buildrel \Delta \over = \frac{1}{N}E{\left\| {\bf{U}} \right\|^2} = \frac{1}{N}\int_{\cal V} {\frac{{{{\left\| {\bf{x}} \right\|}^2}}}{{{\rm{Vol}}\left( {\cal V} \right)}}} d{\bf{x}}$, where ${\bf{U}}$ is a random vector uniformly distributed over ${\cal V}$. The normalized second-order moment of  ${\cal V}$ is defined as $G\left( \cal V  \right) \buildrel \Delta \over = \frac{{{\sigma ^2}\left( \Lambda  \right)}}{{{\rm{Vol}}{{\left( {\cal V} \right)}^{{2 \mathord{\left/
						{\vphantom {2 N}} \right.
						\kern-\nulldelimiterspace} N}}}}}$, which is invariant to the scale of the lattice (i.e., if the generator matrix ${\bf{G}}$ were to be scaled by a non-zero scalar, $G\left( \Lambda  \right)$ would remain the same).  The above definitions of volume and (normalized) second-order moment are not only restricted to Voronoi regions of lattices; they are applicable to any regions over the space of $\mathbb{R}{^N}$. In other words, in place of ${\cal V}$,  for any subset of $S \subset \mathbb{R}{^N}$, we will also make use of the generalized definitions for  the second moment  ${\sigma ^2}(S)$
and the normalized second moment $G(S)$ later in this paper.

If the lattice $\Lambda '$ is a subset of another lattice $ \Lambda $, $\Lambda ' \subset \Lambda $, we say $\Lambda '$ is nested in $ \Lambda $. A pair of lattices $\left( {\Lambda ',\Lambda } \right)$ is called a nested pair if $\Lambda ' \subset \Lambda $, where $\Lambda '$ is called the coarse lattice and $\Lambda$ is called the fine lattice.  For example, $\left( {q{\mathbb{Z}^N},{\mathbb{Z}^N}} \right)$ is a nested pair, where $q$ is a non-zero integer.

We now apply lattices to the coding problem. In this subsection, we consider a point-to-point single-user channel: ${\bf{y}} = \beta {\bf{x}} + {\bf{n}}$, where ${\bf{x}}$ is the vector of the channel symbols,  ${\bf{y}}$ is the vector of the received signals,  $\beta $ is the channel gain, and ${\bf{n}}$ is the vector of i.i.d. real AWGN components with mean-zero and variance-$\sigma _n^2$. All vectors here are length-$N$ vectors. The vector of the channel symbols ${\bf{x}}$ is the codeword for conveying information message ${\bf{m}}$ and it is subject to an average power constraint ${{E{{\left\| {\bf{x}} \right\|}^2}} \mathord{\left/
		{\vphantom {{E{{\left\| {\bf{x}} \right\|}^2}} N}} \right.
		\kern-\nulldelimiterspace} N} \le {P_X}$. The aim is to achieve the channel capacity $C = \left( {{1 \mathord{\left/
			{\vphantom {1 2}} \right.
			\kern-\nulldelimiterspace} 2}} \right){\log _2}\left( {1 + SNR} \right)$, where $SNR = {{{P_X}{\beta ^2}} \mathord{\left/
		{\vphantom {{{P_X}{\beta ^2}} {\sigma _n^2}}} \right.
		\kern-\nulldelimiterspace} {\sigma _n^2}}$. 

For lattice coding, the codeword ${\bf{x}}$  is a point chosen from a specific lattice $\Lambda$ that contains an infinite number of lattice points. We call $\Lambda$  the coding lattice. The power constraint on ${\bf{x}}$  means that only a finite number of the lattice points of $\Lambda$ can be chosen as the codewords. To limit the power of codewords, we can choose a subset of lattice points within a certain region to be the codewords. This region, typically centered on the zero lattice point, is referred to as the \emph{shaping region}. Overall, we need to take into account the following two aspects for designing lattice codes \cite{erez2004achieving}.

\begin{itemize}
	\item  The granularity of the coding lattice $\Lambda$  is represented by its fundamental Voronoi region ${\cal V}$. The volume ${\rm{Vol}}\left( {\cal V} \right)$  determines the inter-codeword Euclidean distance, hence, it determines the decoding error probability.

	\item  The structure of the shaping region determines the power-volume tradeoff, hence, the gap from the channel capacity.    
\end{itemize}

Two key questions are what is a good lattice for coding and what is a proper shaping region that satisfies the power constraint.

To see what is a good lattice for coding, let us first remove the power constraint on the codeword ${\bf{x}}$. In this case, since the transmission power as well as the data rate is infinite, any point of a lattice can be chosen as the codeword. At the receiver side, the maximum likelihood (ML) decoding is employed to search for a lattice point nearest to the received vector. Obviously, the decision regions of the ML decoding are the Voronoi regions and this ML decoder essentially is the lattice quantizer ${Q_\Lambda }\left( \cdot \right)$. The performance of the code is expressed by the decoding error probability $\Pr \left( {\widehat{\bf{m}} \ne {\bf{m}}} \right)$. Since decoding errors occur when the noise vector goes beyond the Voronoi region of the transmitted lattice point, the decoding error probability is $\Pr \left( {\widehat{\bf{m}} \ne {\bf{m}}} \right) = \Pr \left( {{\bf{n}} \notin {\cal V}} \right)$. The decoding error probability is determined by the ``shape'' of $\cal V$, which depends on the lattice being used, and the size of $\cal V$ relative to the noise power, which can be expressed as the volume-to-noise ratio (VNR) given by $\gamma \left( {\Lambda ,\sigma _n^2} \right) \buildrel \Delta \over = {{{\rm{Vol}}{{\left( {\cal V} \right)}^{{2 \mathord{\left/
						{\vphantom {2 N}} \right.
						\kern-\nulldelimiterspace} N}}}} \mathord{\left/
		{\vphantom {{{\rm{Vol}}{{\left( {\cal V} \right)}^{{2 \mathord{\left/
									{\vphantom {2 N}} \right.
									\kern-\nulldelimiterspace} N}}}} {\sigma _n^2}}} \right.
		\kern-\nulldelimiterspace} {\sigma _n^2}}$.  A good lattice code should have low  $\Pr \left( {{\bf{n}} \notin {\cal V}} \right)$
for a fixed $\gamma \left( {\Lambda ,\sigma _n^2} \right)$. According to \cite{erez2004achieving, erez2005lattices}, we have the following result for lattices good for coding.
\\

\noindent
\emph{Goodness of Lattices for Coding}: A sequence of lattices  ${\Lambda ^{\left( N \right)}}$  indexed by their dimension is said to be good for coding if 
\begin{itemize}
	\item[i)] for a target decoding error probability $\Pr \left( {\widehat{\bf{m}} \ne {\bf{m}}} \right)$, where $0 < \Pr \left( {\widehat{\bf{m}} \ne {\bf{m}}} \right) < 1$,  VNR $\gamma \left( {\Lambda ,\sigma _n^2} \right)$ required to achieve the target $\Pr \left( {\widehat{\bf{m}} \ne {\bf{m}}} \right)$ approaches $2\pi e$ as $N$ goes to infinity (i. e., ${\lim _{N \to \infty }}\gamma \left( {\Lambda ,\sigma _n^2} \right) = 2\pi e$); and
	
	\item[ii)] for a fixed  $\gamma \left( {\Lambda ,\sigma _n^2} \right)$, where $\tilde \gamma  > 2\pi e$, the decoding error probability   $\Pr \left( {\widehat{\bf{m}} \ne {\bf{m}}} \right)$ vanishes exponentially in $N$
	
	that is greater than $2\pi e$, $\Pr \left( {\widehat{\bf{m}} \ne {\bf{m}}} \right)$ vanishes exponentially in $N$ (i.e., ${\lim _{N \to \infty }}\Pr \left( {\widehat{\bf{m}} \ne {\bf{m}}} \right) = 0$).
	)
\end{itemize}
Reference \cite{poltyrev1994coding} showed such lattices exist. 
\\

We next examine the power-constraint case to see how to choose a good shaping region. The (normalized) second-order moment is a metric for the average power of a random vector uniformly distributed over a given shaping region, thus it measures how good the shaping region is.  Among all $N$-dimensional bodies of a fixed volume, the body with the minimum normalized second-moment is the N-dimensional sphere, denoted by ${S^{\left( N \right)}}$. The normalized second-order moment of the $N$-dimensional sphere $G\left( {{S^{\left( N \right)}}} \right)$ decreases monotonically with $N$ and approaches the limit  ${1 \mathord{\left/
		{\vphantom {1 {2\pi e}}} \right.
		\kern-\nulldelimiterspace} {2\pi e}}$ as $N \to \infty $ \cite{forney1998modulation, erez2004achieving}. Thus, the optimal choice for the shaping region is the $N$-dimensional sphere denoted by ${S^{\left( N \right)}}$
\cite{forney1998modulation, erez2004achieving}.

To introduce the notion of  \emph{shaping loss}, we consider a simple choice for the shaping region: the $N$-dimensional hypercube. It is well-known that there is a shaping loss between the average powers of hypercube shaping and sphere shaping \cite{forney1998modulation}. The hypercube is the fundamental Voronoi region of the lattice $\mathbb{Z}{^N}$, and  its normalized second-order moment is  $G\left( {{^N}} \right) = {1 \mathord{\left/
		{\vphantom {1 {12}}} \right.
		\kern-\nulldelimiterspace} {12}}$, $\forall N$. Therefore, compared with the optimal sphere shaping, the shaping loss for the hypercube shaping is ${\gamma _s}\left( {{\mathbb{Z}^N}} \right) \buildrel \Delta \over = {{\left( {{1 \mathord{\left/
					{\vphantom {1 {2\pi e}}} \right.
					\kern-\nulldelimiterspace} {2\pi e}}} \right)} \mathord{\left/
		{\vphantom {{\left( {{1 \mathord{\left/
								{\vphantom {1 {2\pi e}}} \right.
								\kern-\nulldelimiterspace} {2\pi e}}} \right)} {G\left( {{^N}} \right)}}} \right.
		\kern-\nulldelimiterspace} {G\left( {{\mathbb{Z}^N}} \right)}} = {{\pi e} \mathord{\left/
		{\vphantom {{\pi e} 6}} \right.
		\kern-\nulldelimiterspace} 6}$
as $N \to \infty $. This  ${{\pi e} \mathord{\left/
		{\vphantom {{\pi e} 6}} \right.
		\kern-\nulldelimiterspace} 6}$
(1.53 dB) shaping loss for the hypercube shaping dose not vanish as the SNR increases, and thus it is an undesired feature, especially in the high SNR regime \cite{forney1998modulation}. 

To maintain the optimalities for both coding and shaping, \cite{de1989some} showed that a spherical lattice code (the intersection of a lattice good for coding with an $N$-dimensional sphere of radius $\sqrt {NP} $) can arbitrarily approach the channel capacity. Although the sphere region can maintain the shaping optimality, such spherical lattice code destroys the linear structure of the original coding lattice $\Lambda $. Moreover, the optimal ML decoding for such spherical lattice code is not the lattice quantization  ${Q_\Lambda }\left( \cdot \right)$  anymore, because the ML decoding regions for codewords are not identical and some are not bounded.  By contrast, in the unconstraint case, the lattice quantization ${Q_\Lambda }\left( \cdot \right)$   used for ML decoding ignores the boundary of the code and preserves the symmetry of the lattice structure in the decoding process: it is much less complex as far as decoding is concerned. For practical implementation, it is desirable to preserve the linear structure of lattices both in the encoding and decoding processes.

\begin{figure}[!t]
	\centering
	\includegraphics[width=3in]{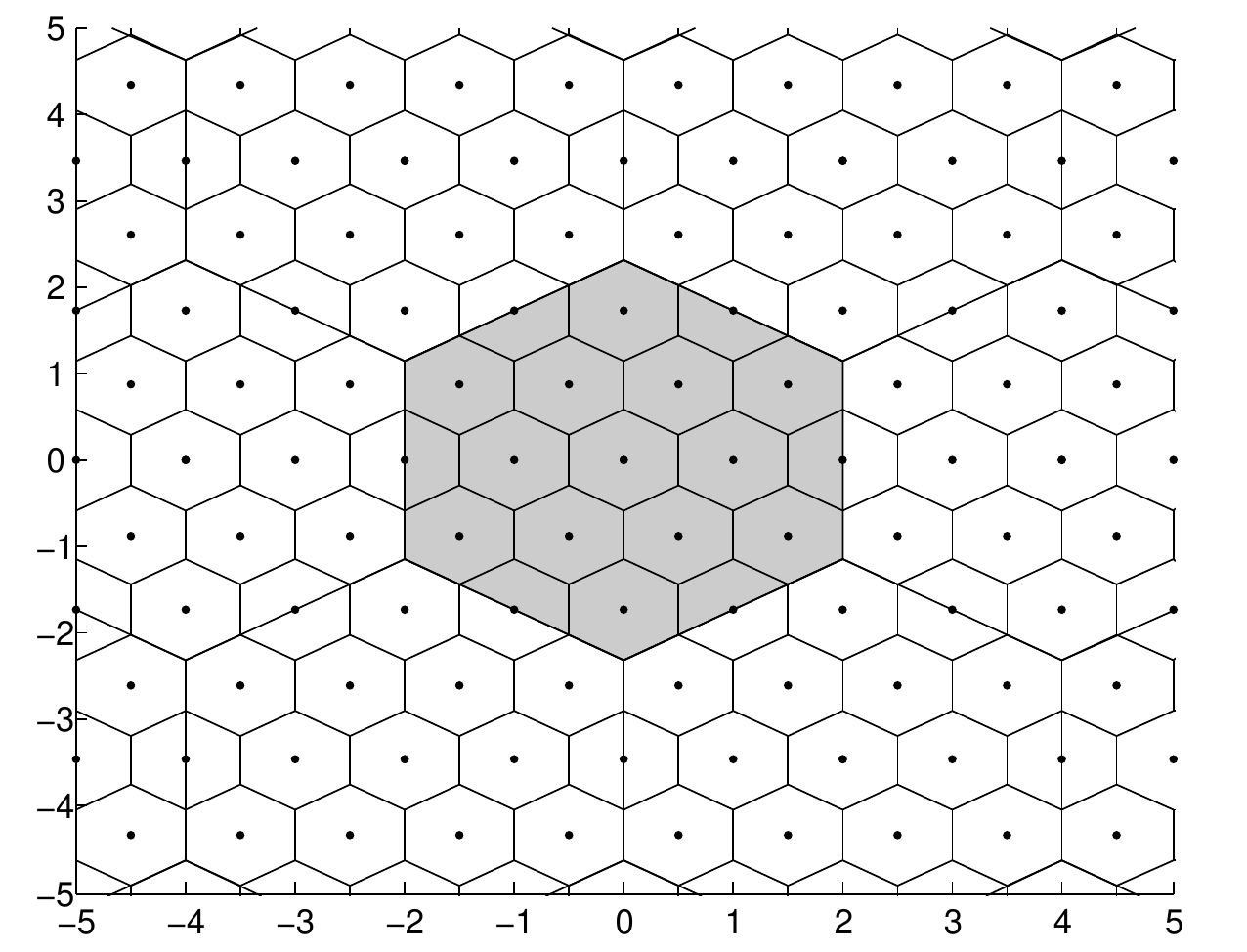}
	\caption{A nested lattice code with $N=2$ and ${\rm{Vol}}\left( {{{\cal V}_s}} \right){\rm{ = 16Vol}}\left( {\cal V} \right)$ (the shaded region is the used codebook).} \label{nested_lattice}
	\vskip -0.1in
\end{figure}

\begin{figure*}[!t]
	\centering
	\includegraphics[width=7in]{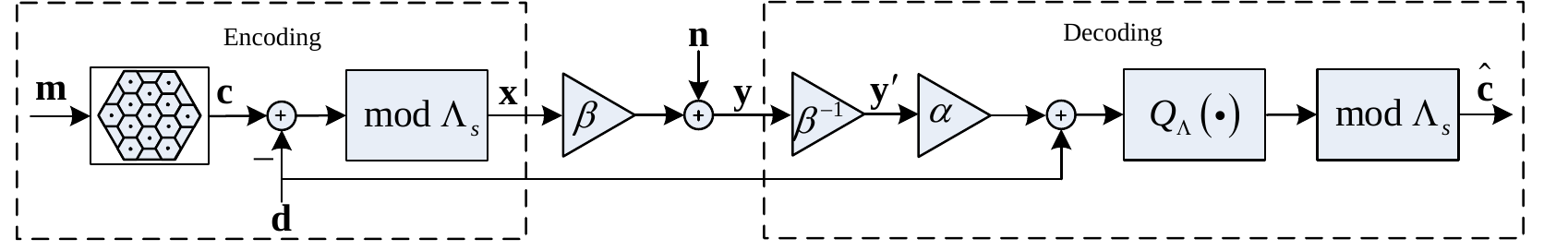}
	\caption{The encoding and decoding processes of nested lattice codes for point-to-point single-user channels.} \label{nested_lattice_p2p}
	\vskip -0.1in
\end{figure*}

Ref. \cite{erez2004achieving} developed a lattice framework that can reliably transmit at rates up to the channel capacity. This framework is called \emph{nested lattice codes} and its general idea is to make use of a nested pair of lattices $\left( {{\Lambda _s},\Lambda } \right)$, where the coarse lattice ${\Lambda _s}$ is used for shaping and the fine lattice ${\Lambda}$ is used for coding. We denote the fundamental Voronoi region of the coarse lattice ${\Lambda _s}$  by ${{\cal V}_s}$, and the volume of ${{\cal V}_s}$ by ${\rm{Vol}}\left( {{{\cal V}_s}} \right)$. The nested lattice code is generated by taking the intersection of the fine lattice used for coding with the fundamental Voronoi region of the coarse lattice used for shaping: ${\cal C} \buildrel \Delta \over = \left\{ {\Lambda  \cap {{\cal V}_s}} \right\}$. The coding rate of the nested lattice code is 
$R = \frac{1}{N}{\log _2}\left| {\cal C} \right| = \frac{1}{N}{\log _2}\frac{{{\rm{Vol}}\left( {{{\cal V}_s}} \right)}}{{{\rm{Vol}}\left( {\cal V} \right)}}.$
Fig.~\ref{nested_lattice} illustrates an example for the codebook of a nested lattice code, where the dimension is $N = 2$ and ${\rm{Vol}}\left( {{{\cal V}_s}} \right){\rm{ = 16Vol}}\left( {\cal V} \right)$ (thus, $R = 2$).  We state the following result for the goodness of lattices for shaping.  
\\

\noindent
\emph{Goodness of Lattices for Shaping}\footnote{In lattice literature, this feature is also termed as the goodness of lattices for MMSE quantization.}: A sequence of lattice  $\Lambda _s^{\left( N \right)}$ is good for shaping if $\mathop {\lim }\limits_{N \to \infty } G\left( {\Lambda _s^{\left( N \right)}} \right) = {1 \mathord{\left/
		{\vphantom {1 {2\pi e}}} \right.
		\kern-\nulldelimiterspace} {2\pi e}}$. Such lattices exist as shown in \cite{zamir1996lattice}. 
\\

It is known that the normalized second-order moment of a lattice is always larger than ${1 \mathord{\left/
		{\vphantom {1 {2\pi e}}} \right.
		\kern-\nulldelimiterspace} {2\pi e}}$, the normalized second-order moment of a sphere with infinite dimensions. The goodness of lattices for shaping indicates that as the dimensions become sufficiently large, there are lattices whose fundamental Voronoi region approach a sphere as their normalized second-order moments go to ${1 \mathord{\left/
		{\vphantom {1 {2\pi e}}} \right.
		\kern-\nulldelimiterspace} {2\pi e}}$. Therefore, such lattice is asymptotically optimal for shaping in terms of that its normalized second-order moment approaches  ${1 \mathord{\left/
		{\vphantom {1 {2\pi e}}} \right.
		\kern-\nulldelimiterspace} {2\pi e}}$. The authors of \cite{erez2005lattices} showed that nested pair of lattices $\left( {{\Lambda _s},\Lambda } \right)$, where the coarse lattice $\Lambda _s$ is good for shaping and the fine lattice $\Lambda$ is good for coding, exist for any required rate. Therefore, based on such nested pair of lattices  $\left( {{\Lambda _s},\Lambda } \right)$, the channel capacity can be potentially be achieved in all SNR regimes. However, before that, there are still two important ingredients of nested lattice codes: the \emph{dithering operation} and \emph{minimum mean square error (MMSE) scaling} at the encoding and decoding processes. We now give the complete description for the encoding and decoding processes of nested lattice codes \cite{erez2004achieving}.    
\\

\noindent
\emph{Encoding and Decoding Processes of Nested Lattice Codes}:
\begin{itemize}
	\item  \emph{Encoding}: First, the message ${\bf{m}}$ is mapped to a codeword ${\bf{c}} = \phi \left( {\bf{m}} \right)$, where $\phi \left( \cdot \right)$ is the message-to-codeword mapping function, the codeword ${\bf{x}}$ belongs to the codebook of the used nested lattice code ${\cal C}  = \left\{ {\Lambda  \cap {{\cal V}_s}} \right\}$   . Then, the transmitted vector is generated according to 
	\begin{equation}\label{encoding}	
		{\bf{x}} = \left[ {{\bf{c}} - {\bf{d}}} \right]{\rm{mod }}~{\Lambda _s}
	\end{equation}	
	where ${\bf{d}} \in {{\cal V}_s}$
	is the dithering vector that is uniformly distributed over the shaping region ${{\cal V}_s}$. The dithering vector ${\bf{d}}$  is known at both of the encoding and decoding processes. 
	
	\item  \emph{Decoding}: The estimate for the transmitted codeword is computed according to 
	\begin{equation}\label{decoding}	
		\widehat{\bf{c}} = {Q_\Lambda }\left( {\left[ {\alpha {\bf{y'}} + {\bf{d}}} \right]} \right){\rm{mod }}~{\Lambda _s}
	\end{equation}	
	where ${\bf{y'}} = {\beta ^{ - 1}}{\bf{y}}$ is the channel-gain-normalized received vector, $\alpha  \buildrel \Delta \over = {{SNR} \mathord{\left/
			{\vphantom {{SNR} {\left( {1 + SNR} \right)}}} \right.
			\kern-\nulldelimiterspace} {\left( {1 + SNR} \right)}}$ is the MMSE coefficient used to scale the channel-gain-normalized received vector ${\bf{y'}}$ before sending it to the lattice quantizer for decoding. 
	\\
	
\end{itemize}

The above encoding and decoding processes of nested lattice codes are illustrated in Fig.~\ref{nested_lattice_p2p}. The dithering operation in (\ref{encoding}) can ensure that the distribution of ${\bf{x}}$ is the same as that of ${\bf{d}}$ (c.f. Lemma 1 of \cite{erez2004achieving}). Therefore, as long as the used shaping lattice is scaled to have the second-order moment $P_X$, the power of the transmitted vector of symbols ${\bf{x}}$ is $P_X$. Furthermore, if the shaping region ${{\cal V}_s}$ approaches a sphere as $N \to \infty $, ${\bf{x}}$ will have white Gaussian distributions, as desired by Shannon theory. The use of MMSE scaling in (\ref{decoding}) plays an important role for the purpose of achieving the channel capacity, especially in the low SNR regime. Please see \cite{erez2004achieving} for more details about the MMSE scaling. In \cite{erez2004achieving}, it was proved that the lattice quantizer decoding in (\ref{decoding}) suffices to be optimal. We end this preliminary on nested lattice codes here. The reader is referred to \cite{erez2004achieving} for further details. In conclusion, with nested lattices codes, the channel capacity $C = \left( {{1 \mathord{\left/
			{\vphantom {1 2}} \right.
			\kern-\nulldelimiterspace} 2}} \right){\log _2}\left( {1 + SNR} \right)$ can be achieved in all SNR regimes. In the next section, employing nested lattice codes, we will present a framework for achieving the capacity pair of the rate-diverse wireless network coding.

\subsection{Nested Lattice Framework for Achieving Capacity Pair}

\begin{figure*}[!t]
	\centering
	\includegraphics[width=7in]{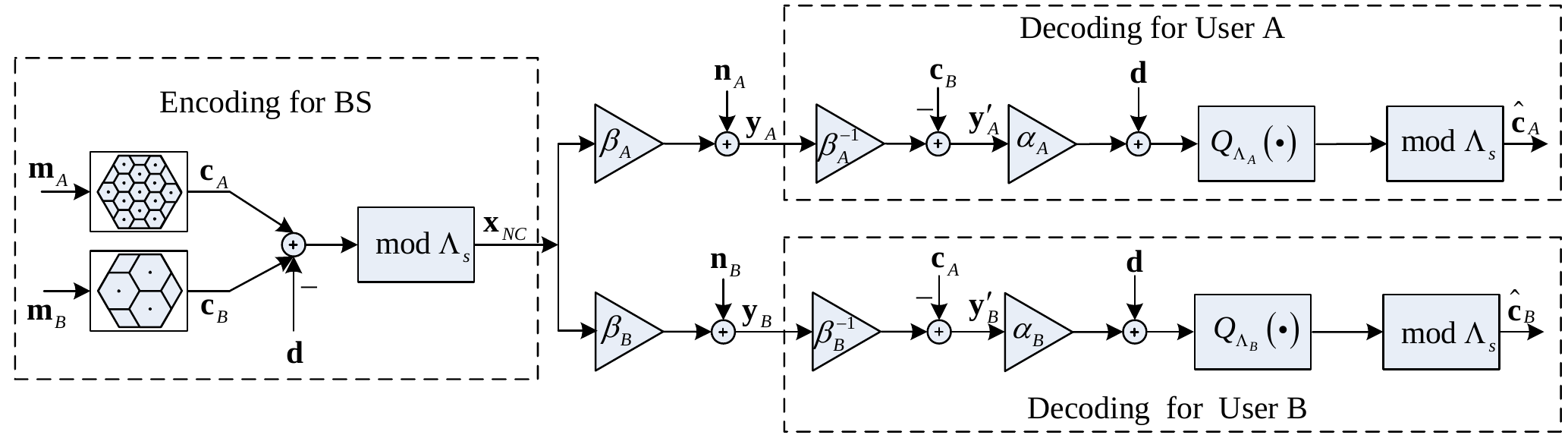}
	\caption{The illustration for the proposed encoding/decoding framework with nested lattice codes for rate-diverse wireless network coding.} \label{nested_lattice_nc}
	\vskip -0.1in
\end{figure*}

Consider the rate-diverse network coding problem. The two users have different channel qualities, thus different channel capacities. As a consequence, the codes operating at different channels have different rates. To develop two nested lattice codes with different rates, we employ two nested pairs of lattices $\left( {{\Lambda _s},{\Lambda _A}} \right)$,  $\left( {{\Lambda _s},{\Lambda _B}} \right)$, where the two different fine lattices ${\Lambda _A}$  and ${\Lambda _B}$ are used for the coding of user A and B respectively, the same coarse lattice ${\Lambda _s}$ is used for the shaping. The corresponding two nested lattice codes are ${{\cal C}_A} = \left\{ {{\Lambda _A} \cap {{\cal V}_s}} \right\}$, ${{\cal C}_B} = \left\{ {{\Lambda _B} \cap {{\cal V}_s}} \right\}$.  We employ ${{\cal C}_A}$, ${{\cal C}_B}$ in our nested lattice framework to achieve the capacity pair of the rate-diverse network coding problem.

The proposed encoding scheme $f\left( \cdot \right)$ at the transmitter of BS first maps the messages ${{\bf{m}}_A}$, ${{\bf{m}}_B}$ into the codewords: ${{\bf{c}}_A} = {\phi _A}\left( {{{\bf{m}}_A}} \right)$, ${{\bf{c}}_B} = {\phi _B}\left( {{{\bf{m}}_B}} \right)$, where ${{\bf{c}}_A} \in {{\cal C}_A}$, ${{\bf{c}}_B} \in {{\cal C}_B}$, and ${\phi _A}\left(  \cdot\right)$, ${\phi _B}\left( \cdot \right)$  are the message-to-codeword mapping functions for codes ${{\cal C}_A}$, ${{\cal C}_B}$. Then, we perform the network coding operation over the codewords to form the network-coded codeword
\begin{equation}\label{nc_encoding}	
	{{\bf{c}}_{NC}} = \left[ {{{\bf{c}}_A} + {{\bf{c}}_B}} \right]{\rm{mod }}~{\Lambda _s}
\end{equation}	
Finally, like nested lattice codes for point-to-point channels, we perform dithering operation to generate the vector of channel symbols 
\begin{equation}\label{nc_dither}	
	{{\bf{x}}_{NC}} = \left[ {{{\bf{c}}_{NC}} - {\bf{d}}} \right]{\rm{mod }}~{\Lambda _s}
\end{equation}
Since ${{\bf{c}}_{NC}} \in {{\cal V}_s}$, the distribution of ${{\bf{x}}_{NC}}$ is still a uniform distribution over the shaping region ${\Lambda _s}$ (the same as the dithering vector $\bf{d}$). Thus, if ${\Lambda _s}$ approaches a sphere as the $N$ grows, the transmitted vector ${{\bf{x}}_{NC}}$  will look like a white Gaussian noise. This satisfies the requirement on the channel inputs by Shannon theory. Using the distributive law of the modulo arithmetic, the operations (\ref{nc_encoding}) and (\ref{nc_dither}) can be combined into 
\begin{equation}\label{lattice_nc}
	\begin{array}{l}
		{{\bf{x}}_{NC}} = \left[ {\left[ {{{\bf{c}}_A} + {{\bf{c}}_B}} \right]{\rm{mod}}~{\Lambda _s} - {\bf{d}}} \right]{\rm{mod}}~{\Lambda _s} \\ 
		\;\;\;\;\;\;\;\;= \left[ {{{\bf{c}}_A} + {{\bf{c}}_B} - {\bf{d}}} \right]{\rm{mod}}~{\Lambda _s} \\ 
	\end{array}
\end{equation}
Based on the expression in (\ref{lattice_nc}), we illustrate the proposed encoding scheme for rate-diverse wireless network coding in Fig.~\ref{nested_lattice_nc}.

The decoding process at the receiver of user A, ${g_A}\left(  \cdot  \right)$, performs the following steps in sequence: the network decoding (the substruction of its side-information), the scaling and de-dithering, and the lattice quantization with respect to the coding lattice ${\Lambda _A}$. Thus, the estimate for the target message of user A is given by 
\begin{equation}\label{lattice_nc_decoding}
	{\widehat{\bf{c}}_A} = \underbrace {{Q_{{\Lambda _A}}}\left( {\underbrace {\left[ {{\alpha _A}\underbrace {\left( {\beta _A^{ - 1}{{\bf{y}}_A} - {{\bf{c}}_B}} \right)}_{{\rm{Step~I}}\;:{\rm{Network}}\:{\rm{Decoding}}} + {\bf{d}}} \right]}_{{\rm{Step~II}}\;:{\rm{Scaling~ and ~Dedithering}}}} \right)}_{{\rm{Step~III}}\;:{\rm{Lattice}}\:{\rm{Quantization}}}{\rm{mod}}~{\Lambda _s}
\end{equation}
where ${\alpha _A} \buildrel \Delta \over = {{SN{R_A}} \mathord{\left/
		{\vphantom {{SN{R_A}} {\left( {1 + SN{R_A}} \right)}}} \right.
		\kern-\nulldelimiterspace} {\left( {1 + SN{R_A}} \right)}}$
is the MMSE coefficient for user A. User B performs a similar decoding process. The decoding processes for rate-diverse wireless network coding are also illustrated in Fig.~\ref{nested_lattice_nc}.

Substituting the received vector at the user A, ${{\bf{y}}_A} = {\beta _A}{{\bf{x}}_{NC}} + {{\bf{n}}_A}$, into (\ref{lattice_nc_decoding}) and making some manipulations give 
\begin{equation}\label{lattice_nc_decoding2}
	\begin{array}{l}
		{\widehat{\bf{c}}_A} = {Q_{{\Lambda _A}}}\left( {\left[ {{\alpha _A}{{\bf{c}}_A} - {\alpha _A}{\bf{d}} + {\alpha _A}\beta _A^{ - 1}{{\bf{n}}_A} + {\bf{d}}} \right]} \right){\rm{mod }}{\Lambda _s} \\ 
		= {Q_{{\Lambda _A}}}\left( {\left[ {{\alpha _A}{{\bf{c}}_A} - {\alpha _A}{\bf{d}} + {\alpha _A}\beta _A^{ - 1}{{\bf{n}}_A} + {\bf{d}}} \right]{\rm{mod }}{\Lambda _s}} \right) \\ 
		= {Q_{{\Lambda _A}}}\left( {\left[ {{\alpha _A}\left( {\underbrace {\underbrace {\left[ {{{\bf{c}}_A} - {\bf{d}}} \right]{\rm{mod }}{\Lambda _s}}_{ \buildrel \Delta \over = {{\bf{x}}_A}} + \beta _A^{ - 1}{{\bf{n}}_A}}_{ \buildrel \Delta \over = {{{\bf{y'}}}_A}}} \right) + {\bf{d}}} \right]{\rm{mod }}{\Lambda _s}} \right) \\ 
		= {Q_{{\Lambda _A}}}\left( {\left[ {{\alpha _A}{{{\bf{y'}}}_A} + {\bf{d}}} \right]{\rm{mod }}{\Lambda _s}} \right) \\ 
		= {Q_{{\Lambda _A}}}\left( {\left[ {{\alpha _A}{{{\bf{y'}}}_A} + {\bf{d}}} \right]} \right){\rm{mod }}{\Lambda _s} \\ 
	\end{array}
\end{equation}
where ${{\bf{x}}_A} \buildrel \Delta \over = \left[ {{{\bf{c}}_A} - {\bf{d}}} \right]{\rm{mod }}~{\Lambda _s}$ that is the vector of virtual point-to-point channel symbols obtained by performing single-user encoding on message ${{\bf{m}}_A}$ using nested lattice code ${{\cal C}_A}$ designed for user A, and ${{\bf{y'}}_A} \buildrel \Delta \over = {{\bf{x}}_A} + \beta _A^{ - 1}{{\bf{n}}_A}$ is the equivalent channel model for user A after the network decoding operation,   

In (\ref{lattice_nc_decoding2}), we can see that the equivalent channel left for user A, ${{\bf{y'}}_A} = {{\bf{x}}_A} + \beta _A^{ - 1}{{\bf{n}}_A}$, is the same as a point-to-point single-user channel where message ${{\bf{m}}_A}$ is conveyed by vector ${{\bf{x}}_A}$. We also note that the decoding operation for rate-diverse wireless network coding expressed by (\ref{lattice_nc_decoding2}) has the same form as the decoding operation in a point-to-point single-user channel expressed by (\ref{decoding}). Compared to the point-to-point single-user channel from BS to user A, the SNR of the equivalent channel is still $SN{R_A} = {{{P_X}\beta _A^2} \mathord{\left/
		{\vphantom {{{P_X}\beta _A^2} {\sigma _n^2}}} \right.
		\kern-\nulldelimiterspace} {\sigma _n^2}}$ (i.e. the SNR is not reduced in the rate-diverse wireless network coding case). In this sense, user B is totally transparent to user A.  Therefore, if the used nested pair $\left( {{\Lambda _s},{\Lambda _A}} \right)$ is good for shaping and coding for the point-to-point single-user channel of user A, we can achieve the channel capacity ${C_A}$ using nested lattice code ${\cal{C}_A}$. With the same decoding scheme, the same result holds for user B. Therefore, as long as nested pairs $\left( {{\Lambda _s},{\Lambda _A}} \right)$, $\left( {{\Lambda _s},{\Lambda _B}} \right)$ are both good for shaping and coding, the above proposed encoding/decoding framework can achieve the capacity pair $\left( {{C_A},{C_B}} \right)$. Such nested pairs do exist and can be obtained using the construction method proposed in \cite{nazer2011compute}.

With the linear structured nested lattice codes, the above encoding/decoding framework is optimal for achieving the capacity pair of the rate-diverse network coding. The framework brings out a general design principle for the encoding/decoding processes. The principle, referred to as \emph{the principle of virtual single-user channels}, can be stated as follows:

\noindent \textbf{Principle of Virtual Single-User Channels} :

\begin{itemize}
	\item[i)] At the transmitter side, the encoding scheme performs coding for the two users separately, using the codes with different rates that matches with the capacities of their respective channels. The two codewords are then combined by a network coding operation. 
	\item[ii)] At the receiver side, each user applies network decoding on the received signal so as to extract the signal containing only the desired codeword. The signal of the other user should be totally transparent to the user after the network decoding process. In particular, the remaining signal is exactly the same as that of a single-user point-to-point communication, including the effective signal strength and the effective noise thereof. Applying single-user decoding to extract the desired message has performance exactly the same as that in conventional point-to-point communication.  
\end{itemize}

The above nested-lattice-code framework adheres to this design principle. As will be seen later, this principle can also be realized using other codes with simpler decoding algorithms. 

Although the nested-lattice-code framework is optimal theoretically, its exact implementation faces many difficulties. In particular, the lattice quantization decoding, which searches the transmitted lattice point over the lattice space, has unaffordable complexity as the codeword length $N$ grows. For the existing lattice quantization  decoding methods, the complexity increases exponentially with $N$. There have been efforts to design practical lattice codes with implementable decoding algorithms \cite{LDLC2008, shalvi2011signal}. In the next section, we employ one of such codes, LDLC  \cite{LDLC2008}, to realize the design principle. In the section after that, we show that the same design principle can also be realized by BICM.

\section{Implementation of the Proposed Framework Using LDLC}

We provide the background on LDLC in Section IV.A. We apply LDLC in our rate-diverse wireless network coding framework in Section IV.B.

\subsection{Low Density Lattice Codes (LDLC)}

As discussed in Section III, lattice codes use the points of a specific lattice as codewords (messages). The codewords of a lattice code are generated by ${\bf{x}} = {\bf{Gb}}$, where ${\bf{x}} \in \mathbb{R} {^N}$ is the codeword, ${\bf{G}} \in {\mathbb{R}^{N \times N}}$ is the generating matrix\footnote{We assume square LDLC generating matrices in this section. The results can be  easily extended to non-square generating matrices.}, and ${\bf{b}} \in \mathbb{Z} {^N}$ is the integer message vector.  LDLC is a type of lattice codes where the parity check matrix, defined as the inverse of the generating matrix ${\bf{H}} \buildrel \Delta \over = {{\bf{G}}^{ - 1}}$,  is \emph{sparse} \cite{LDLC2008}. 

The number of non-zero elements in the $i$-th row (column) of ${\bf{H}}$ is called the degree of the $i$-th row (column). The sparse structure of ${\bf{H}}$  implies that the degrees of its all rows and columns are far less than $N$. For LDLC, the non-zero elements of ${\bf{H}}$ can adopt values in $\mathbb{R}$, in contrast to  ordinary LDPC where the elements of ${\bf{H}}$  are typically elements of a finite field.

In practice, the elements of the integer message vector $\bf{b}$ only can be selected from some finite constellations, rather than the infinite $\mathbb{Z}^{N}$. We denote the $i$-th element of the integer message vector $\bf{b}$ by ${b_i} \in {{\cal L}_i}$, where ${{\cal L}_i}$  is a subset of integers. The coding rate of such LDLC is $R = {{\sum\nolimits_{i = 1}^N {{{\log }_2}\left| {{{\cal L}_i}} \right|} } \mathord{\left/
		{\vphantom {{\sum\nolimits_{i = 1}^N {{{\log }_2}\left| {{{\cal L}_i}} \right|} } N}} \right.
		\kern-\nulldelimiterspace} N}$, where $\left| {{{\cal L}_i}} \right|$   is the cardinality of  ${{\cal L}_i}$. 

In \cite{LDLC2008}, Latin square LDLC were considered. Latin square LDLC are defined by imposing some restrictions to their parity check matrixes. For a Latin square LDLC, all rows and columns of  ${\bf{H}}$ have the same degree $d$;  a sorted sequence of $d$  non-zero values, $\left\{ {{h_d} \ge {h_{d - 1}} \ge  \cdots  \ge {h_1} > 0} \right\}$, is the generating sequence of ${\bf{H}}$; the $d$ non-zero elements in one row or column of ${\bf{H}}$  are $\left\{ {{h_d},{h_{d - 1}}, \cdots ,{h_1}} \right\}$ with a change of positions and signs for different rows and columns. Given a generating sequence, the parity check matrix of a LDLC can be constructed using the proposed method in \cite{LDLC2008}. For example, $\left\{ {1,0.8,0.5} \right\}$  is a generating sequence and 
$${\bf{H}} = \left( {\matrix{
		0 & { - 0.8} & 0 & { - 0.5} & 1 & 0  \cr 
		{0.8} & 0 & 0 & 1 & 0 & { - 0.5}  \cr 
		0 & {0.5} & 1 & 0 & {0.8} & 0  \cr 
		0 & 0 & { - 0.5} & { - 0.8} & 0 & 1  \cr 
		1 & 0 & 0 & 0 & {0.5} & {0.8}  \cr 
		{0.5} & { - 1} & { - 0.8} & 0 & 0 & 0  \cr 		
	} } \right)$$
	is the corresponding parity matrix of a LDLC with degree $d=3$ and dimension $N=6$.

	\begin{figure*}[!t]
		\centering
		\begin{minipage}[t]{.32\linewidth}
			\subfigure[]{
				\includegraphics[width=2.4in]{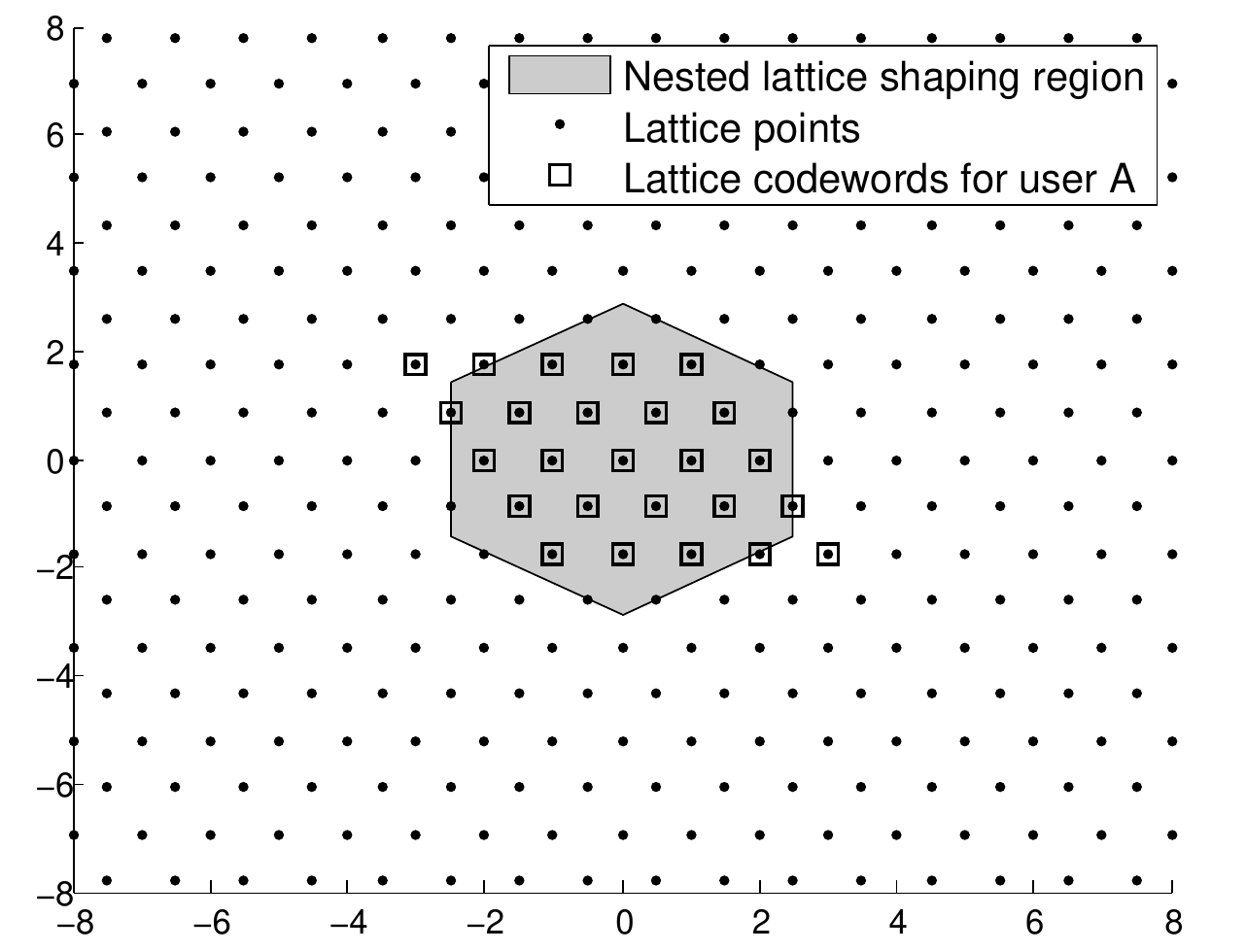}}

		\end{minipage}
		\begin{minipage}[t]{.32\linewidth}
			\subfigure[]{
				\includegraphics[width=2.4in]{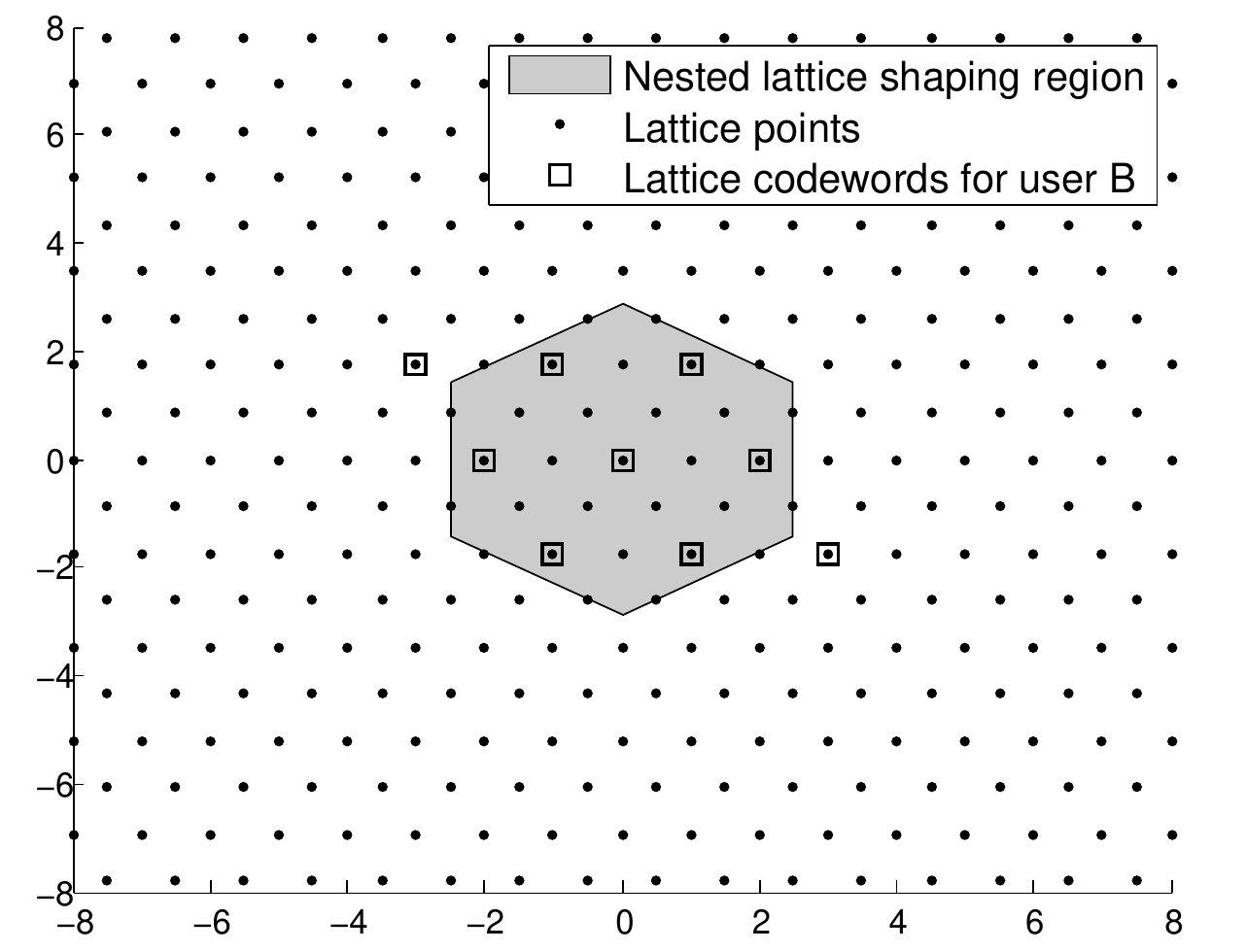} }

		\end{minipage}
		\begin{minipage}[t]{.32\linewidth}
			\subfigure[]{
				\includegraphics[width=2.4in]{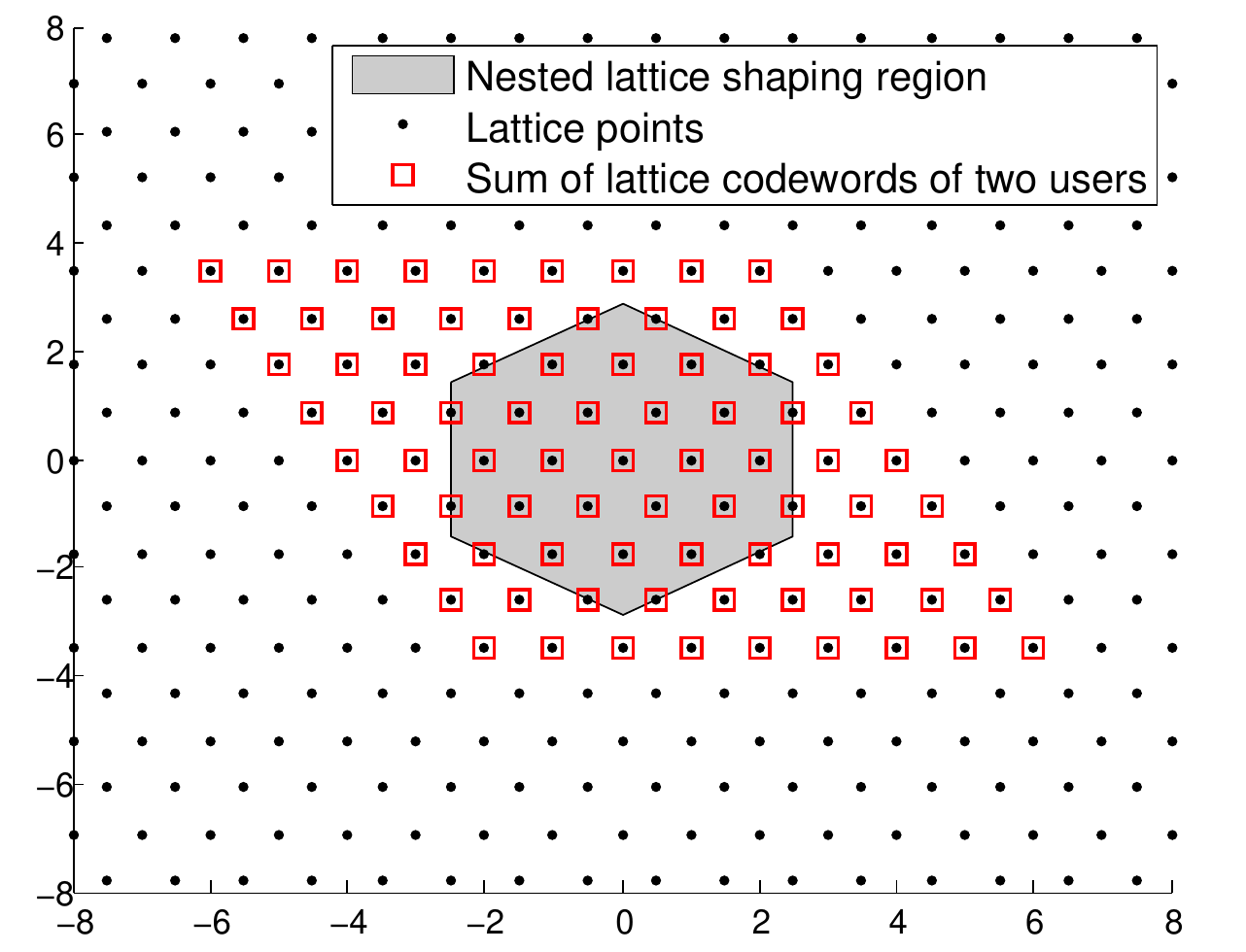} }
		\end{minipage}
		
		\caption{An example of the coding and shaping aspects for single-user LDLC and wireless network coding LDLC: (a) Single-user LDLC with coding rate 2.32; (b) Single-user LDLC with coding rate 1.59; (c) Wireless network coding LDLC obtained by directly adding together the lattice codewords of user A in (a) and that of user B in (b).} \label{ldlc}
		
		\vskip -0.1in
	\end{figure*}

	For communications over power-constrained AWGN channels, only codewords within some shaping region can be chosen as the valid codewords. Directly transmitting codewords of LDLC without shaping will increase the average transmission power. Fig.~\ref{ldlc} (a) illustrates this situation using an example. In Fig~\ref{ldlc} (a), we generate a two-dimensional lattice using the generating matrix ${\bf{G}} = \left( {\matrix{
			1 & { - 1/2}  \cr 
			0 & {{{\sqrt 3 } \mathord{\left/
						{\vphantom {{\sqrt 3 } 2}} \right.
						\kern-\nulldelimiterspace} 2}}  \cr 
		} } \right)$.  The black dots are the lattice points. The squares are the lattice codewords selected by ${\bf{x}} = {\bf{Gb}}$, where ${{\cal L}_i} = \left\{ { - 2, - 1,0,1,2} \right\}$for all $i$. We have $25$ lattice points are selected as the codewords, and the coding rate thus is $R = {{{{\log }_2}25} \mathord{\left/
		{\vphantom {{{{\log }_2}25} 2}} \right.
		\kern-\nulldelimiterspace} 2} = 2.3$. Fig.~\ref{ldlc} (a) also shows the nested lattice shaping region. From Fig.~\ref{ldlc} (a), we can intuitively see that the codewords outside the shaping region will greatly increase the average transmission power.

To reduce the average transmission power, we must map all the codewords ${\bf{x}} = {\bf{Gb}}$ into the codewords ${\bf{x'}}$ that are within the shaping region. This shaping operation can be expressed by mapping integer message vectors $\bf b$  to some other integer vectors ${\bf{b'}}$,  such that the obtained  codewords ${\bf{x'}} = {\bf{Gb'}}$   lie within the shaping region. Note that ${b_i}^\prime$ may not be an element of ${{\cal L}_i}$ anymore. We will describe how to perform shaping for LDLC in next subsection.

Due to the sparse structure of ${\bf{H}}$, a belief propagation (BP) algorithm can be used to decode the LDLC codeword at the receiver side. Given the parity check matrix ${\bf{H}}$, we can construct the factor graph that models the constraint imposed by LDLC encoding. Then, we derive the BP decoding as a message passing algorithm on the factor graph using the sum-product rule \cite{kschischang2001factor}. Since the factor graph for LDLC contains loops, the BP algorithm will need to go through several iterations. Although the BP decoding algorithm is in general not optimal for lattice decoding, its linear complexity is its advantage over more complicated decoding in practice. Compared with BP decoding for LDPC, BP decoding for LDLC computes continuous-valued messages rather than discrete-valued message; the check nodes of the factor graph force the summation of their incoming messages to integers with certain pre-defined range. For details on BP decoding for LDLC, we refer the readers to \cite{LDLC2008}.

\subsection{Wireless Network Coding Using LDLC}

We now apply LDLC to the rate-diverse wireless network coding problem. The codewords ${{\bf{c}}_A}$, ${{\bf{c}}_B}$ transmitted to user A and B are generated by applying LDLC encoding: ${{\bf{c}}_A} = {{\bf{G}}_A}{{\bf{b}}_A}$, ${{\bf{c}}_B} = {{\bf{G}}_B}{{\bf{b}}_B}$, where ${{\bf{b}}_u}$ is the integer message vector of user $u$, ${{\bf{G}}_u}$ is the LDLC generating matrix of user $u$.  After network coding and lattice shaping, the transmitted codeword is ${\bf c}_u$. Note that the dithering vector ${\bf{d}}$ in (6) usually is not used in real systems, and thus we do not consider it hereafter. 

The lattices ${\Lambda _A}$ and ${\Lambda _B}$ generated by ${{\bf{G}}_A}$  and ${{\bf{G}}_B}$ are the coding lattice for user A and B, respectively. For shaping purposes, our optimal framework requires a coarse lattice ${\Lambda _s}$  that is simultaneously nested in the two coding lattices, i.e., ${\Lambda _s} \subset {\Lambda _A}$, ${\Lambda _s} \subset {\Lambda _B}$. The code rate for user A (B) is determined by the intersection between ${\Lambda _A}$ (${\Lambda _B}$) and ${{\cal V}_s}$.  The inverse matrices of  ${{\bf{G}}_A}$  and  ${{\bf{G}}_B}$ must be sparse to allow practical BP decoding. The generating matrices  ${{\bf{G}}_A}$, ${{\bf{G}}_B}$, ${{\bf{G}}_s}$  of such  ${\Lambda _A}$, ${\Lambda _B}$, ${\Lambda _s}$   can be found as follows.

We first find an LDLC generating matrix ${\bf{G}}$ using the construction method proposed in \cite{LDLC2008}. The inversion of ${\bf{G}}$ is sparse. We call ${\bf{G}}$ the basic generating matrix. Different coding rates are assigned to the two users by applying different constellations to the integer message vectors of the two users, i.e., ${b_{u,i}} \in {{\cal L}_{u,i}} \buildrel \Delta \over = \left\{ { - {L_{u,i}}, - {L_{u,i}} + 1, \cdots ,{L_{u,i}} - 1} \right\}$, where integer ${b_{u,i}}$ is the $i$-th element of ${{\bf{b}}_u}$, ${{\cal L}_{u,i}}$ is the constellation for ${b_{u,i}}$, and ${L_{u,i}}$ is an integer parameter that determines the corresponding constellation size via $\left| {{{\cal L}_{u,i}}} \right| = 2{L_{u,i}}$. Note that, for the same user u, we may assign different constellations to different integer message element ${b_{u,i}}$. To limit the average transmission power, we choose the shaping lattice as ${{\bf{G}}_s} = {\bf{GM}}$, where ${\bf{M}} \buildrel \Delta \over = diag\left\{ {{M_1},{M_2}, \cdots ,{M_N}} \right\}$ is an $N \times N$ diagonal matrix. The $i$-th main diagonal element of ${\bf{M}}$, $M_i$, is set to be the least common multiple (LCM) of $2{L_{A,i}}$ and $2{L_{B,i}}$: ${M_i} \buildrel \Delta \over = LCM\left\{ {2{L_{A,i}},2{L_{B,i}}} \right\}$. Finally, we choose the coding lattice for user $u$ as ${{\bf{G}}_u} = {\bf{G}}{{\bf{M}}_u}$, where ${{\bf{M}}_u} \buildrel \Delta \over = diag\left\{ {{M_{u,1}},{M_{u,2}}, \cdots ,{M_{u,N}}} \right\}$ is an $N \times N$ diagonal integer matrix whose $i$-th main diagonal element is set as ${M_{u,i}} \buildrel \Delta \over = {{{M_i}} \mathord{\left/
		{\vphantom {{{M_i}} {2{L_{u,i}}}}} \right.
		\kern-\nulldelimiterspace} {2{L_{u,i}}}}$. With the above setup, we can easily check that ${\Lambda _s} \subset {\Lambda _A}$, ${\Lambda _s} \subset {\Lambda _B}$ are satisfied --- the shaping optimality of the nested lattice framework for rate-diverse wireless network coding is guaranteed. 

Fig.~\ref{ldlc} (a) and Fig.~\ref{ldlc} (b) show examples for LDLC lattices with two dimensions ($N=2$) for users A and B, respectively. The black dots are the lattice points generated by the basic generating matrix ${\bf{G}}$. The squares in Fig.~\ref{ldlc} (a) are the lattice codewords for user A selected by applying constellation ${{\cal L}_{A,i}} = \left\{ { - 2, - 1,0,1,2} \right\}$ to ${b_{A,i}}$ for all $i$. The squares in Fig.~\ref{ldlc} (b) are the lattice codewords for user B selected by applying constellation ${\mathcal{L}_{B,i}} = \left\{ { - 1,0,1} \right\}$ to ${b_{B,i}}$  for all $i$. The code rates for users A and B are 2.32 and 1.59 respectively. In Fig.~\ref{ldlc} (c), the red squares indicate the lattice codewords obtained by directly adding together the lattice codewords of user A in Fig.~\ref{ldlc} (a) and that of user B in Fig.~\ref{ldlc} (b) without shaping. From Fig.~\ref{ldlc} (c), we can see that, without shaping, the average transmission power increases quite heavily compared with the single-user case, due to the addition. Therefore, shaping is more crucial for the network coding case. The next question is how to perform the nested lattice shaping for rate-diverse wireless network coding using the coarse lattice ${\Lambda _s}$.

As long as the coarse lattice  ${\Lambda _s}$ (equivalently, ${{\bf{G}}_s}$) used for shaping is specified, the lattice shaping operation, ${{\bf{x}}_{NC}} = \left[ {{{\bf{c}}_A} + {{\bf{c}}_B}} \right]{\rm{mod }}{\Lambda _s}$, can be expressed as 
\begin{equation}\label{ldlc_shap1}
	\begin{array}{l}
		{{\bf{x}}_{NC}} = {{\bf{c}}_A} + {{\bf{c}}_B} - {Q_{{\Lambda _s}}}\left( {{{\bf{c}}_A} + {{\bf{c}}_B}} \right) \\ 
		\;\;\;\;\;\;\;\; = {{\bf{c}}_A} + {{\bf{c}}_B} - {{\bf{G}}_s}{\bf{k}} \\ 
	\end{array}
\end{equation}
where ${{\bf{G}}_s}{\bf{k}} = {Q_{{\Lambda _s}}}\left( {{{\bf{c}}_A} + {{\bf{c}}_B}} \right)$, ${\bf{k}} \in {\mathbb{Z}^N}$, is the coarse lattice point that is closest to ${{\bf{c}}_A} + {{\bf{c}}_B}$. Substituting ${{\bf{c}}_A} = {{\bf{G}}_A}{{\bf{b}}_A} = {\bf{G}}{{\bf{M}}_A}{{\bf{b}}_A}$, ${{\bf{c}}_A} = {{\bf{G}}_B}{{\bf{b}}_B} = {\bf{G}}{{\bf{M}}_B}{{\bf{b}}_B}$ and ${{\bf{G}}_s} = {\bf{GM}}$ into (\ref{ldlc_shap1}), we have  
\begin{equation}\label{ldlc_shap2}
	\begin{array}{l}
		{{\bf{x}}_{NC}} = {\bf{G}}{{\bf{M}}_A}{{\bf{b}}_A} + {\bf{G}}{{\bf{M}}_B}{{\bf{b}}_B} - {\bf{GMk}} \\ 
		= {\bf{G}}\left( {\underbrace {\underbrace {{{\bf{M}}_A}{{\bf{b}}_A} + {{\bf{M}}_B}{{\bf{b}}_B}}_{ \buildrel \Delta \over = {{\bf{b}}_{NC}}} - {\bf{Mk}}}_{ \buildrel \Delta \over = {{{\bf{b'}}}_{NC}}}} \right) \\ 
		= {\bf{G}}{{{\bf{b'}}}_{NC}} \\ 
	\end{array}
\end{equation}
where ${{\bf{b'}}_{NC}} \buildrel \Delta \over = {{\bf{b}}_{NC}} - {\bf{Mk}}$ with ${{\bf{b}}_{NC}} \buildrel \Delta \over = {{\bf{M}}_A}{{\bf{b}}_A} + {{\bf{M}}_B}{{\bf{b}}_B}$. The shaping operation in (\ref{ldlc_shap2}) is to map the network-coded integer message vector ${{\bf{b}}_{NC}}$ to another network-coded integer message vector ${{\bf{b'}}_{NC}} = {{\bf{b}}_{NC}} - {\bf{Mk}}$,  such that the resulting coding lattice point (the codeword ${{\bf{x}}_{NC}} = {\bf{G}}{{\bf{b'}}_{NC}}$) lies within the fundamental Voronoi region of the shaping lattice.

The target of the shaping operation is to solve the quantization problem ${{\bf{G}}_s}{\bf{k}} = {Q_{{\Lambda _s}}}\left( {{{\bf{c}}_A} + {{\bf{c}}_B}} \right)$, which is equivalent to finding the integer vector 
\begin{equation}\label{findk}
	{\bf{k}} = \arg \mathop {\min }\limits_{{\bf{k'}}} {\left\| {{{\bf{c}}_A} + {{\bf{c}}_B} - {{\bf{G}}_s}{\bf{k'}}} \right\|^2}
\end{equation}
Since the inverse of ${{\bf{G}}_s} = {\bf{GM}}$ is still sparse, at first sight, we can also employ the BP algorithm developed for decoding LDLC to solve the shaping problem (\ref{findk}) approximately. However, \cite{LDLC2008} showed that the BP algorithm cannot give satisfactory shaping performance. The reason is that the distribution of the “effective noise” (i.e., the deviation from ${{\bf{G}}_s}{\bf{k}}$) in the shaping problem (i.e., the codeword ${{\bf{c}}_A} + {{\bf{c}}_B}$) is a uniform over the space of codewords, not a Gaussian distribution such as that of channel noise. Due to the uniform distribution, the probability of the codewords ${{\bf{c}}_A} + {{\bf{c}}_B}$  appearing at the boundaries of the Voronoi regions of the shaping lattice will not decrease with the distance between the codewords and the centers of the Voronoi regions. Therefore, the variance of the effective noise in the shaping problem is much larger than that in the decoding problem. The BP algorithm that iteratively and approximately solves the decoding problem fails to reduce the power of  ${{\bf{x}}_{NC}}$ as a result. 

Alternatively, \cite{sommer2009shaping} employed the M-algorithm to solve the shaping problem for LDLC in point-to-point channels. For the shaping problem of LDLC in rate-diverse wireless network coding, we also use the M-algorithm.

The M-algorithm is a sequential tree search algorithm \cite{aulin1999breadth}. To apply it to the shaping problem (\ref{findk}), we first perform QR decomposition on the LDLC parity check matrix ${{\bf{H}}^T} = {{\bf{Q}}^T}{\bf{R}}$, where ${{\bf{Q}}^T}$ is an orthogonal matrix and  ${\bf{R}}$ is a upper triangle matrix. Then, ${\bf{H}}$ can be written as ${\bf{H}} = {\bf{TQ}}$, where ${\bf{T}} = {{\bf{R}}^T}$ is a lower triangle matrix. Multiplying ${\bf{H}} = {\bf{TQ}}$ on both sides of ${{\bf{x}}_{NC}} = {\bf{G}}{{\bf{b'}}_{NC}}$, we have the following new check relationship:
\begin{equation}\label{ma1}
	{\bf{T}}{\widetilde{\bf{x}}_{NC}} = {{\bf{b'}}_{NC}} = {{\bf{b}}_{NC}} - {\bf{Mk}}
\end{equation}
where ${\widetilde{\bf{x}}_{NC}} \buildrel \Delta \over = {\bf{Q}}{{\bf{x}}_{NC}}$ is the codeword after the transformation by ${\bf{Q}}$. Since ${\bf{T}}$ is a lower triangle matrix, the check relationship (\ref{ma1}) can be decomposed into $N$ check equations
\begin{equation}\label{ma2}
	\begin{array}{r}
		\sum\limits_{j = 1}^{i - 1} {{T_{i,j}}{{\widetilde x}_{NC,j}}}  + {T_{i,i}}{\widetilde x_{NC,i}} = {b_{NC,i}} - {M_i}{k_i},{\rm{    }} \\ 
		i = 1,2, \cdots ,N \\ 
	\end{array}
\end{equation}
where ${T_{i,j}}$ is the ${\left( {i,j} \right)}$-th element of ${\bf{T}}$. The $i$-th check equation in (\ref{ma2}) corresponds to the $i$-th row of (\ref{ma1}). Originally, the shaping problem in (\ref{findk}) aims to minimize ${\left\| {{{\bf{x}}_{NC}}} \right\|^2}$. Since ${\widetilde{\bf{x}}_{NC}} = {\bf{Q}}{{\bf{x}}_{NC}}$ and ${\bf{Q}}$ is an orthogonal matrix that will not change the power of ${{\bf{x}}_{NC}}$, we now change our objective to minimizing ${\left\| {{{\widetilde{\bf{x}}}_{NC}}} \right\|^2}$ to solve the shaping problem. The triangular structure of ${\bf{T}}$  in (\ref{ma1}) suggests that we can make use of a tree search method, such as the M-algorithm, over all possible vector ${\bf{k}}$ to find the one that minimize the power of  ${\widetilde{\bf{x}}_{NC}}$. 

We consider a tree rooted at a dummy node. The nodes at the $i$-th level of the tree are labeled by different sequences of ${k_1},{k_2}, \cdots ,{k_i}$. We note that given a particular sequence ${k_1},{k_2}, \cdots ,{k_i}$, the sequence ${\widetilde x_{NC,1}},{\widetilde x_{NC,2}}, \cdots ,{\widetilde x_{NC,i}}$ is also fixed due to (\ref{ma2}). Each node ${k_1},{k_2}, \cdots ,{k_i}$ is associated with a metric $\sum\nolimits_{j = 1}^i {{{\left\| {{{\widetilde x}_{NC,j}}} \right\|}^2}} $.  The optimal ML tree search requires tracing all the paths on the tree. The number of possible paths on the tree increases exponentially with the depth of the level. Since the depth of the tree, the length of ${\bf{k}}$, is  large for non-trivial lattice codes, this may incur huge computational complexities. Therefore, sub-optimal algorithms with lower complexity, but with good performance, are needed to perform the tree search task for LDLC shaping. The sub-optimal M-algorithm \cite{aulin1999breadth} starts the tree search at the root node, and visits the levels from the one to the next one. At every level of the tree, the M-algorithm just retains $M$ paths rather than all the possible paths, where $M$  is a design parameter. The node at the $i$-th  level ${k_1},{k_2}, \cdots ,{k_{i - 1}},{k_i}$ is a child node of its parent node ${k_1},{k_2}, \cdots ,{k_{i - 1}}$ at the $\left( {i - 1} \right)$-th level, and the metric of the node ${k_1},{k_2}, \cdots ,{k_{i - 1}},{k_i}$ is recursively computed as
\begin{equation}
	\sum\limits_{j = 1}^i {{{\left\| {{{\widetilde x}_{NC,j}}} \right\|}^2}}  = \sum\limits_{j = 1}^{i - 1} {{{\left\| {{{\widetilde x}_{NC,j}}} \right\|}^2}}  + {\left\| {{{\widetilde x}_{NC,i}}} \right\|^2}
\end{equation}
where 
\begin{equation}
	{\widetilde x_{NC,i}} = \frac{1}{{{T_{i,i}}}}\left( {{b_{NC,i}} - \sum\limits_{j = 1}^{i - 1} {{T_{i,j}}{{\widetilde x}_{NC,j}} - {M_i}{k_i}} } \right)
\end{equation}
is the $i$-th element of ${\widetilde{\bf{x}}_{NC}}$ computed according to (\ref{ma2}). We assume ${k_i} \in \left\{ { - \left\lceil {{{{M_i}} \mathord{\left/
				{\vphantom {{{M_i}} 2}} \right.
				\kern-\nulldelimiterspace} 2}} \right\rceil , - \left\lceil {{{{M_i}} \mathord{\left/
				{\vphantom {{{M_i}} 2}} \right.
				\kern-\nulldelimiterspace} 2}} \right\rceil  + 1, \cdots ,\left\lfloor {{{{M_i}} \mathord{\left/
				{\vphantom {{{M_i}} 2}} \right.
				\kern-\nulldelimiterspace} 2}} \right\rfloor  - 1} \right\}$.  Among all the paths reached at the $i$-th  level, we just retain the $M$ paths that have the least $M$ values of the metric $\sum\nolimits_{j = 1}^i {{{\left\| {{{\widetilde x}_{NC,j}}} \right\|}^2}} $.  Then, the search continues to the $\left( {i + 1} \right)$-th level of the tree. After the search finishes visiting the leaf nodes of the tree, we obtain $M$ paths labeled by the sequence  ${k_1},{k_2}, \cdots ,{k_N}$, and finally we select the one with the least  metric value $\sum\nolimits_{j = 1}^N {{{\left\| {{{\widetilde x}_{NC,j}}} \right\|}^2}} $ as the final estimate for ${\bf{k}}$.

At the receiver of user A, a network decoding operation first substracts the side information from the received signal: ${{\bf{y'}}_A} = {{\bf{y}}_A} - {\beta _A}{\bf{G}}{{\bf{M}}_B}{{\bf{b}}_B}$. Then, BP channel decoding is used to estimate ${{\bf{b'}}_A} = {{\bf{M}}_A}{{\bf{b}}_A} - {\bf{Mk}}$. After that, the estimate for the target message is immediately obtained by the modulo operation ${\widehat{\bf{b}}_A} = {\bf{M}}_A^{ - 1}\left( {{{{\bf{b'}}}_A}{\rm{mod }}~{\bf{M}}} \right)$, which can be performed in an element-by-element manner: ${\widehat b_{A,i}} = M_{A,i}^{ - 1}\left( {{{b'}_{A,i}}{\rm{mod }}~{M_i}} \right)$ for all $i$. We remark here that the use of the modulo operation for recovering ${{\bf{b}}_A}$ (${{\bf{b}}_A}$) avoids the need to derive ${\bf{k}}$ directly at the receivers.

We remark here that the original constellation of ${b_{u,i}}$ is ${{\cal L}_{u,i}} = \left\{ { - {L_{u,i}}, - {L_{u,i}} + 1, \cdots ,{L_{u,i}} - 1} \right\}$ ; after shaping, the constellation of ${b'_{u,i}}$ is changed to ${\mathcal{L}_{u,i}} + {M_i}{k_i}$. For the decoding of the unshaped LDLC, the receiver just considers the possible values of  ${b_{u,i}}$ over ${\mathcal{L}_{u,i}}$. For the decoding of the shaped LDLC, the receiver needs to know the constellation of ${b'_{u,i}}$. Here, we can empirically choose a constellation for ${b'_{u,i}}$. Typically, for the shaping operation, the possible values of ${k_i}$ is bounded by $\left| {{k_i}} \right| \le \varepsilon $ for all $i$, where    $\varepsilon $ is a small positive integer. For example, in our simulations of Section VI, we observed that $\varepsilon  = 2$. Therefore, we can expand the constellation of  ${b'_{u,i}}$ to ${{\cal L}'_{u,i}} = \left\{ { - {L_{u,i}} - \varepsilon {M_i}, - {L_{u,i}} - \varepsilon {M_i} + 1, \cdots ,{L_{u,i}}+\varepsilon {M_i} - 1} \right\}$ and use it in the decoding.\footnote{ In practice, the transmitter of BS can embed $\varepsilon $ in the packet being sent as metadata to facilitate the decoding by receiver.}

Since the principle of virtual single-user channels is satisfied, we predict that the performance of the above LDLC based rate-diverse wireless network coding scheme is the same as that of a LDLC based single-user point-to-point scheme. Moreover, as will be elaborated in the next section, there is an additional shaping gain for the user with the lower data rate.

\subsection{Additional Shaping Gain for Rate-Diverse Wireless Network Coding Using LDLC}

In the optimal lattice coding framework, the shaping operation employs a dithering vector   ${\bf{d}}$ to disseminate the transmission power over the whole shaping region, as indicated by (\ref{encoding}) for the point-to-point case and by (\ref{lattice_nc}) for the network coding case. The distribution of the dithering vector is a continuous uniform distribution. The distribution of lattice points is a discrete uniform distribution. The average power of the lattice points within a shaping region can be changed by varying the density of these lattice points. By increasing the density of the lattice points within the shaping region (piling more lattice points into the shaping region), the average power can approach its lower bound --- the average power of a continuous and uniform random variable over the shaping region. After the shaping operation as in (\ref{encoding}), the distribution of the codeword ${\bf{x}}$ is the same as the dithering vector ${\bf{d}}$ (a continuous and uniform random variable). Therefore, from a theoretical perspective, this dithering operation can reduce the transmission power of a nested lattice code. However, in practice, this dithering vector is not used in the coding process of real systems, since the generation and storage of  the  $N$-dimensional random vector ${\bf{d}} \in \mathbb{R}{^N}$ will impose a large complexity on the system as $N$ grows.

Without dithering, we perform the shaping operation for network coding as ${{\bf{x}}_{NC}} = \left[ {{{\bf{c}}_A} + {{\bf{c}}_B}} \right]{\rm{mod }}~{\Lambda _s}$.  Due to the network coding operation, the lattice points of user A (B) behave somewhat like the dithering vector for user B (A). The density of the codewords, ${{\bf{x}}_{NC}}$, will be the higher one between the densities of ${{\bf{c}}_A}$ and ${{\bf{c}}_B}$. Therefore, compared with the point-to-point case, the user with a lower data rate (lower density of its lattice points within the shaping region) achieves an additional shaping gain. We can  demonstrate this additional shaping gain mathematically using a simple one-dimensional lattice  example.

Consider two one-dimensional lattices: ${\Lambda _A} = \mathbb{Z}$ for user A and ${\Lambda _B} = 2\mathbb{Z}$ for user B.  The shaping region is taken as $\left[ { - L,L} \right]$, where we assume $L$ is a position even number for simplicity. As a consequence, the  codewords of the nested lattice codes for users A and B are ${c_A} \in \left\{ { - L, - L + 1, \cdots ,L - 1,L} \right\}$ and ${c_B} \in \left\{ { - L, - L + 2, \cdots ,L - 2,L} \right\}$ respectively.  The average powers of the codes for user A and B are ${P_A} = E\left( {c_A^2} \right) = {{L\left( {L + 1} \right)} \mathord{\left/
		{\vphantom {{L\left( {L + 1} \right)} 3}} \right.
		\kern-\nulldelimiterspace} 3}$ and ${P_B} = E\left( {c_B^2} \right) = {{L\left( {L + 2} \right)} \mathord{\left/
		{\vphantom {{L\left( {L + 2} \right)} 3}} \right.
		\kern-\nulldelimiterspace} 3}$. After the operations of network coding and  shaping: ${x_{NC}} = \left[ {{c_A} + {c_B}} \right]{\rm{mod }}~2L$, the transmitted codewords are ${x_{NC}} \in \left\{ { - L, - L + 1, \cdots ,L - 1,L} \right\}$ and the average power of ${x_{NC}}$  is ${P_{NC}} = E\left( {x_{NC}^2} \right) = {P_A} = {{L\left( {L + 1} \right)} \mathord{\left/
		{\vphantom {{L\left( {L + 1} \right)} 3}} \right.
		\kern-\nulldelimiterspace} 3}$.  This means that, compared with the point-to-point case, there is an additional shaping gain, ${{10{{\log }_{10}}{P_B}} \mathord{\left/
		{\vphantom {{10{{\log }_{10}}{P_B}} {{P_{NC}}}}} \right.
		\kern-\nulldelimiterspace} {{P_{NC}}}} = 10{\log _{10}}{{\left( {L + 2} \right)} \mathord{\left/
		{\vphantom {{\left( {L + 2} \right)} {\left( {L + 1} \right)}}} \right.
		\kern-\nulldelimiterspace} {\left( {L + 1} \right)}}$, for user B, which has a lower data rate in the rate-diverse wireless network coding case.  We can observe this additional shaping gain from the numerical results in Section VI. 

Recall that our design principle of virtual single-user channels states that we should strive for a design which achieves the single-user performance. The implication is that single-user performance is some sort of an upper bound to be aimed for, and that the two-user rate-diverse system cannot have performance better than the best of single-user systems. The above result, at first sight, may appear to have contradicted the above statement.  A more careful examination can resolve this ``paradox'' easily. Note that the additional shaping gain in the two-user rate-diverse system is caused by the denser effective codebook, which induces some sort of a dithering effect on an un-dithered system. If dithering had been used, then the denser codebook would have no effect. In other words, it would not possible for the two-user rate-diverse system to beat the single-user system if dithering had been applied in the latter in the first place (i.e., it is not possible to beat the best single-user system).

\section{Implementation of the Proposed Framework Using BICM}

BICM is a framework for combining binary channel codes with high-order modulations (e. g., QAM) \cite{caire1998bit}. For the single-user point-to-point problem, BICM first encodes the binary information bits using a binary code, then interleaves the coded bits, and then maps the interleaved bits to modulated signals.. By varying the code rate of the binary channel code and the size of the modulation signal set, BICM can support various target data rates. Moreover, it has been shown that the achievable rate of BICM is very close to the single-user channel capacity \cite{caire1998bit,forney1998modulation}.

\begin{figure*}[!t]
	\centering
	\includegraphics[width=6in]{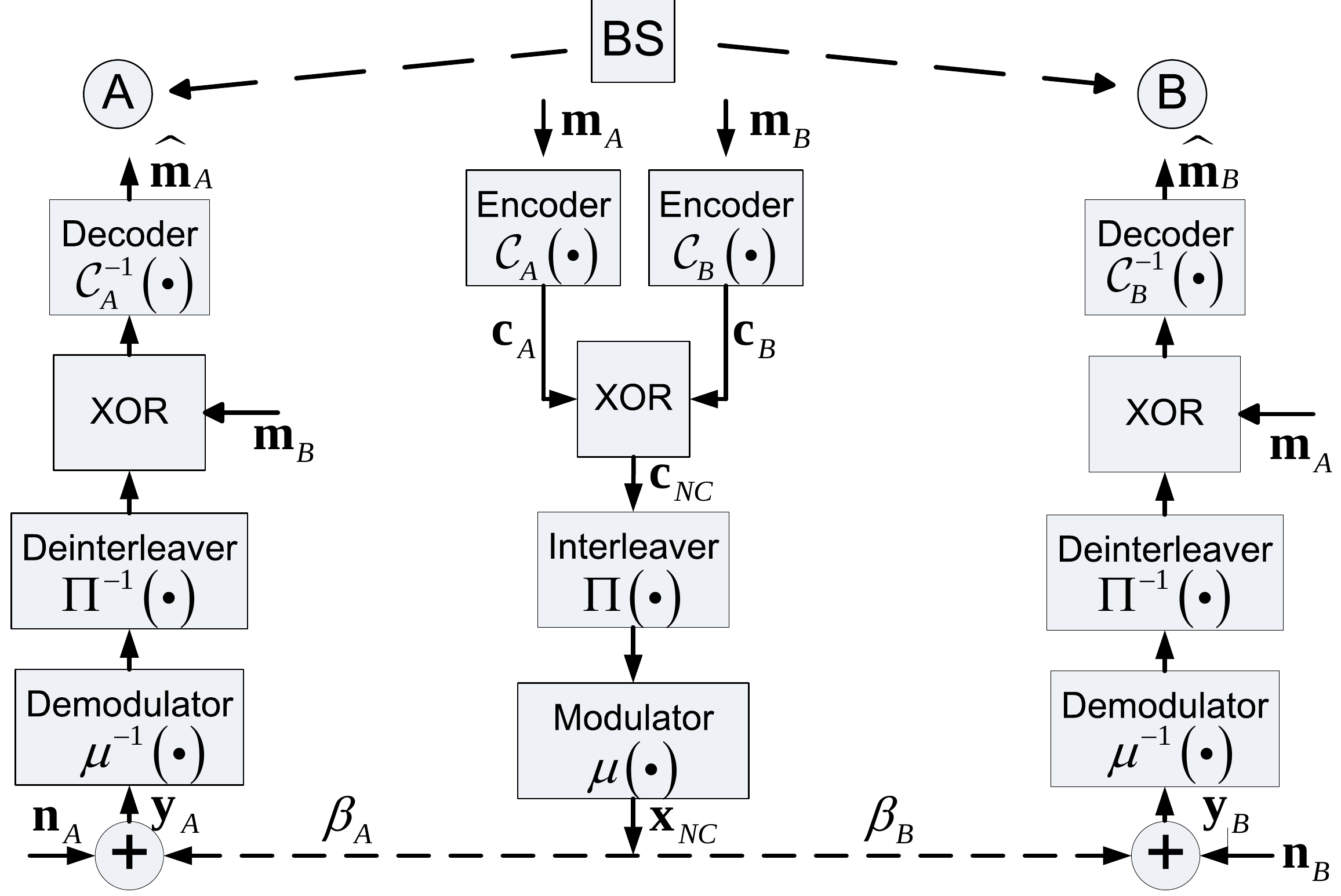}
	\caption{The block diagram of our BICM scheme for rate-diverse wireless network coding.} \label{bicm_system_model}
	
	\vskip -0.2in
	
\end{figure*}

We next show how to apply BICM to the rate-diverse wireless network coding problem. We design the BICM system to adhere to \emph{the principle of virtual single-user channels} expounded in Section III.B. Fig.~\ref{bicm_system_model} shows the block diagram of our proposed BICM scheme. The BS first encodes the binary information bits of users A and B (${{\bf{m}}_A} \in {\left\{ {0,1} \right\}^{{L_A}}}$, ${{\bf{m}}_B} \in {\left\{ {0,1} \right\}^{{L_B}}}$) into binary coded bits: ${{\bf{c}}_A} = {{\cal C}_A}\left( {{{\bf{m}}_A}} \right)$, ${{\bf{c}}_B} = {{\cal C}_B}\left( {{{\bf{m}}_B}} \right)$, where ${{\cal C}_u}\left(  \cdot  \right)$ is the binary channel code for user $u$,  ${{\bf{c}}_u}$ is a length-$L$ vector containing the coded bits of user $u$. The code rates of  ${{\bf{c}}_A}$ and ${{\bf{c}}_B}$  are ${r_A} = {{{L_A}} \mathord{\left/
		{\vphantom {{{L_A}} L}} \right.
		\kern-\nulldelimiterspace} L}$ and ${r_B} = {{{L_B}} \mathord{\left/
		{\vphantom {{{L_B}} L}} \right.
		\kern-\nulldelimiterspace} L}$, respectively.  Next, we perform bit-wise XOR operations over the elements of ${{\bf{c}}_A}$ and ${{\bf{c}}_B}$ as the network coding operation: ${{\bf{c}}_{NC}} = {{\bf{c}}_A} \oplus {{\bf{c}}_B}$. We then interleave the network-coded bits ${{\bf{c}}_{NC}}$ using an interleaver $\pi $: ${\widetilde{\bf{c}}_{NC}} = \pi \left( {{{\bf{c}}_{NC}}} \right)$, and lastly we map the interleaved bits ${\widetilde{\bf{c}}_{NC}}$ to the modulated signals.  

We assume the use of QAM, and its signal set is denoted by ${\cal X}$.  The size of ${\cal X}$ is $\left| {\cal X} \right| = {2^m}$, where $m$ is a positive integer. The signals in ${\cal X}$ are represented by complex numbers. The interleaved bits ${\widetilde{\bf{c}}_{NC}}$ are divided into ${{N \buildrel \Delta \over = L} \mathord{\left/
		{\vphantom {{N \buildrel \Delta \over = L} m}} \right.
		\kern-\nulldelimiterspace} m}$ groups, with each group containing $m$ bits. The   $i$-th group is denoted by ${\widetilde{\bf{c}}_{NC,i}} = {\left[ {{{\widetilde c}_{NC,(i - 1)m + 1}},{{\widetilde c}_{NC,(i - 1)m + 2,}} \cdots ,{{\widetilde c}_{NC,im}}} \right]^T}$, where ${\widetilde c_{NC,j}}$ is the $j$-th element of ${\widetilde{\bf{c}}_{NC}}$, and $i = 1,2, \cdots ,N$. Each group of $m$ interleaved bits is mapped to a modulated signal in ${\cal X}$ via a one-to-one labeling map $\mu :{\left\{ {0,1} \right\}^m} \to {\cal X}$. The $i$-th group ${\widetilde{\bf{c}}_{NC,i}}$ is mapped to the modulated signal ${x_{NC,i}} \in \cal{X}$. We refer to ${\widetilde{\bf{c}}_{NC,i}}$ as the label of ${x_{NC,i}}$. We use the same Gray labeling map $\mu$  as that used in single-user point-to-point BICM schemes \cite{caire1998bit}. The modulated signals ${x_{NC,i}}$, $i = 1,2, \cdots ,N$, are stacked into a vector  ${{\bf{x}}_{NC}} \buildrel \Delta \over = {\left[ {{x_{NC,1}},{x_{NC,2}}, \cdots ,{x_{NC,N}}} \right]^T}$ for transmission to the users over the broadcast channel.

The received signal at a user’s receiver is given by  ${{\bf{y}}_u} = {\beta _u}{{\bf{x}}_{NC}} + {{\bf{n}}_u}$.  We denote the $i$-th element of ${{\bf{y}}_{NC}}$ by  ${y_{NC,i}}$, the subset of the signals ${x_{NC,i}} \in {\cal X}$ where the $j$-th bit of the label of ${x_{NC,i}}$ (${\widetilde c_{NC,(i - 1)m + j}}$)  has the value $b \in \left\{ {0,1} \right\}$ by ${\cal X}_j^{\left( b \right)}$, for $j = 1,2, \cdots ,m$, and $i = 1,2, \cdots ,N$.  Based on ${y_{NC,i}}$, the demodulator computes the detection metric of  ${\widetilde c_{NC,(i - 1)m + j}}$: 
\begin{equation}\label{bicm_metric}
	\begin{array}{l}
		\lambda \left( {{y_{u,i}},{{\widetilde c}_{NC,(i - 1)m + j}} = b} \right) \\ 
		= \sum\limits_{{x_{NC,i}} \in {\cal X}_j^{\left( b \right)}} {p\left( {{y_{u,i}}\left| {{x_{NC,i}}} \right.} \right)p\left( {{x_{NC,i}}\left| {{{\widetilde c}_{NC,(i - 1)m + j}} = b} \right.} \right)}  \\ = \frac{1}{{{2^{m - 1}}}}\sum\limits_{{x_{NC,i}} \in {\cal X}_j^{\left( b \right)}} {p\left( {{y_{u,i}}\left| {{x_{NC,i}}} \right.} \right)}  \\ 
	\end{array}
\end{equation}
for $b \in \left\{ {0,1} \right\}$, $j = 1,2, \cdots ,m$, and $i = 1,2, \cdots ,N$. Then, these detection metrics are forwarded to the deinterleaver ${\pi ^{ - 1}}$,  the output of which are the detection metrics of the network-coded bits $\left\{ {\lambda \left( {{y_{u,i}},{c_{NC,l}} = b} \right)} \right\}$, where $b \in \left\{ {0,1} \right\}$,    and $l = 1,2, \cdots ,L$ is the $l$-th bit of ${\widetilde{\bf{c}}_{NC}}$. Using side information, users can easily deduce the detection  metrics of  the channel-coded bits of its target message. For example, user A obtains the detection metric of ${c_{A,l}}$ by using the side information ${c_{B,l}}$ as follows:
\begin{equation}\label{bicm_dnc}
	\lambda \left( {{y_{A,i}},{c_{A,l}} = b} \right) = \lambda \left( {{y_{A,i}},{c_{NC,l}} = b \oplus {c_{B,l}}} \right)
\end{equation}
where  $b \in \left\{ {0,1} \right\}$, $l = 1,2, \cdots ,L$. Finally, these detection metrics of channel-coded bits are input  to the channel decoder for recovery of the target information bits.  Some remarks about the above BICM scheme are in order:

\begin{itemize}
	\item The BICM scheme for rate-diverse wireless network coding adheres to \emph{the principle of virtual single-user channels} expounded in Section III.  Given the side information, the two users are actually served by two equivalent single-user point-to-point BICM schemes over two independent channels. Thus, single-user decoding after network decoding is sufficient to achieve the performance of the corresponding single-user point-to-point BICM scheme without losing optimality.

	\item The data rates of users A and B are ${R_A} = m{r_A}$, ${R_B} = m{r_B}$. By employing different channel code rates ${r_A}$, ${r_B}$, the equivalent single-user point-to-point BICM schemes can flexibly support different data rates for user A and B. According to the point-to-point channel capacities of the users, ${{\cal C}_A}$ and ${{\cal C}_B}$, we can tune the values of ${r_A}$, ${r_B}$ and $m$ to optimize the system performance. In practice, the channel codes with different code rates ${r_A}$, ${r_B}$  can be obtained from a mother code using methods such as puncturing \cite{hagenauer1988rate} and splitting \cite{joo2009design}. 
	
	\item Unlike our scheme here, \cite{chen2010novel} and \cite{yun2010rate} made use of joint modulation for rate-diverse wireless network coding.  Specifically, without considering channel coding, \cite{chen2010novel, yun2010rate} performed the network coding operations as joint modulations where the information bits of the two users are jointly mapped to a constellation point. Using their different side information, the receivers of the two users can equivalently observe two sub-constellations with different set sizes. Thus, these joint modulations use two different constellations to deal with the different data rates of the two users. However, compared with the conventional constellations with the same cardinalities for single-user point-to-point channels, the minimum Euclidean distances between the points of the sub-constellations are shortened (See next section for details). This means that the equivalent single-user point-to-point SNRs are reduced. Therefore, the joint modulation schemes in \cite{chen2010novel, yun2010rate} do not conform to our principle of virtual single-user channels that guarantees two equivalent single-user point-to-point channels without any SNR reduction.  By simply varying the code rates of the channel codes, our BICM scheme do not lose  optimality, thanks to the principle of virtual single-user channels. The simulation results in the next section demonstrate the performance advantage of our BICM scheme over the joint modulation schemes in \cite{chen2010novel, yun2010rate}.
	
\end{itemize}

\section{Numerical Results }

This section validates the performance of our schemes via simulations. 

\subsection{The simulations of LDLC Rate-Diverse Wireless Network Coding}

This subsection presents simulation results on the performance of the LDLC implementation of the optimal framework for rate-diverse wireless network coding in section IV. In the simulation, the LDLC generating matrix with degree $d = 7$ is obtained from the generating sequence $\left\{ {1,\frac{1}{{\sqrt 7 }},\frac{1}{{\sqrt 7 }},\frac{1}{{\sqrt 7 }},\frac{1}{{\sqrt 7 }},\frac{1}{{\sqrt 7 }},\frac{1}{{\sqrt 7 }}} \right\}$ using the construction algorithm in \cite{LDLC2008}. The codeword length is $100$. For user A, the constellation of the elements of the integer message vectors is ${{\cal L}_{A,i}} = \left\{ { - 4, - 3, \cdots ,2,3} \right\}$ for all $i$; thus the data rate of user A is ${R_A} = {\log _2}\left| {{{\cal L}_i}} \right| = 3$ bits per channel use. For user B, the constellation of the elements of the integer message vectors is ${{\cal L}_{B,i}} = \left\{ { - 2, - 1,0,1} \right\}$ for all $i$; thus the data rate of user B is ${R_B} = {\log _2}\left| {{{\cal L}_i}} \right| = 2$
bits per channel use. The shaping method is the nested shaping method for LDLC. The BP decoding algorithm is used at the receivers of users A and B. Strictly speaking, the messages (the probability functions) in the BP decoding algorithm have real-valued variables. As in \cite{LDLC2008}, we choose to use discrete vectors to approximate the messages. Specifically, the $i$-th received signal is ${y_i} = {x_i} + {n_i}$; the message associated with the variable node ${x_i}$ is a continuous function $\mu \left( {{x_i}} \right)$. We quantize ${x_i}$  with a quantization step and over a range ${x_i} \in \left[ {{y_i} - {\Delta  \mathord{\left/
			{\vphantom {\Delta  2}} \right.
			\kern-\nulldelimiterspace} 2},{y_i} + {\Delta  \mathord{\left/
			{\vphantom {\Delta  2}} \right.
			\kern-\nulldelimiterspace} 2}} \right]$, where $\Delta $ is the magnitude of the range. In this manner, the message $\mu \left( {{x_i}} \right)$ is represented by a vector. In our simulations, we use a quantization step  ${1 \mathord{\left/
		{\vphantom {1 {128}}} \right.
		\kern-\nulldelimiterspace} {128}}$ over a range of $8$ in magnitude. Thus, each message is represented by a length-1024 vector. The BP decoding performs 100 iterations between the variable nodes and the check nodes of the LDLC factor graph to ensure convergence of the decoding process.

\begin{figure}[!t]
	\centering
	\includegraphics[width=3in]{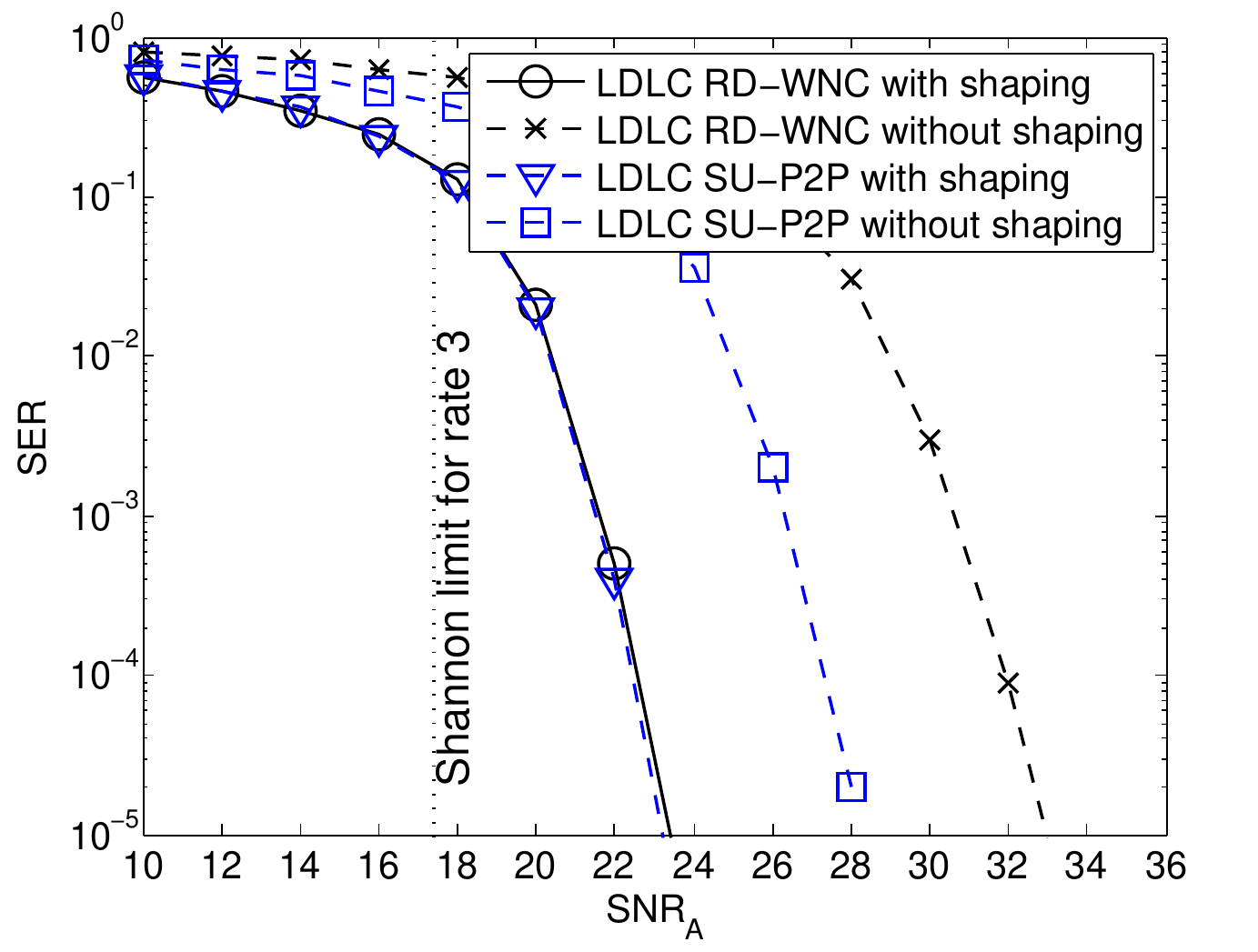}
	\caption{The SER results of LDLC RD-WCN and LDLC SU-P2P for user A.} \label{ldlc_ua}
	\vskip -0.1in
\end{figure}

\begin{figure}[!t]
	\centering
	\includegraphics[width=3in]{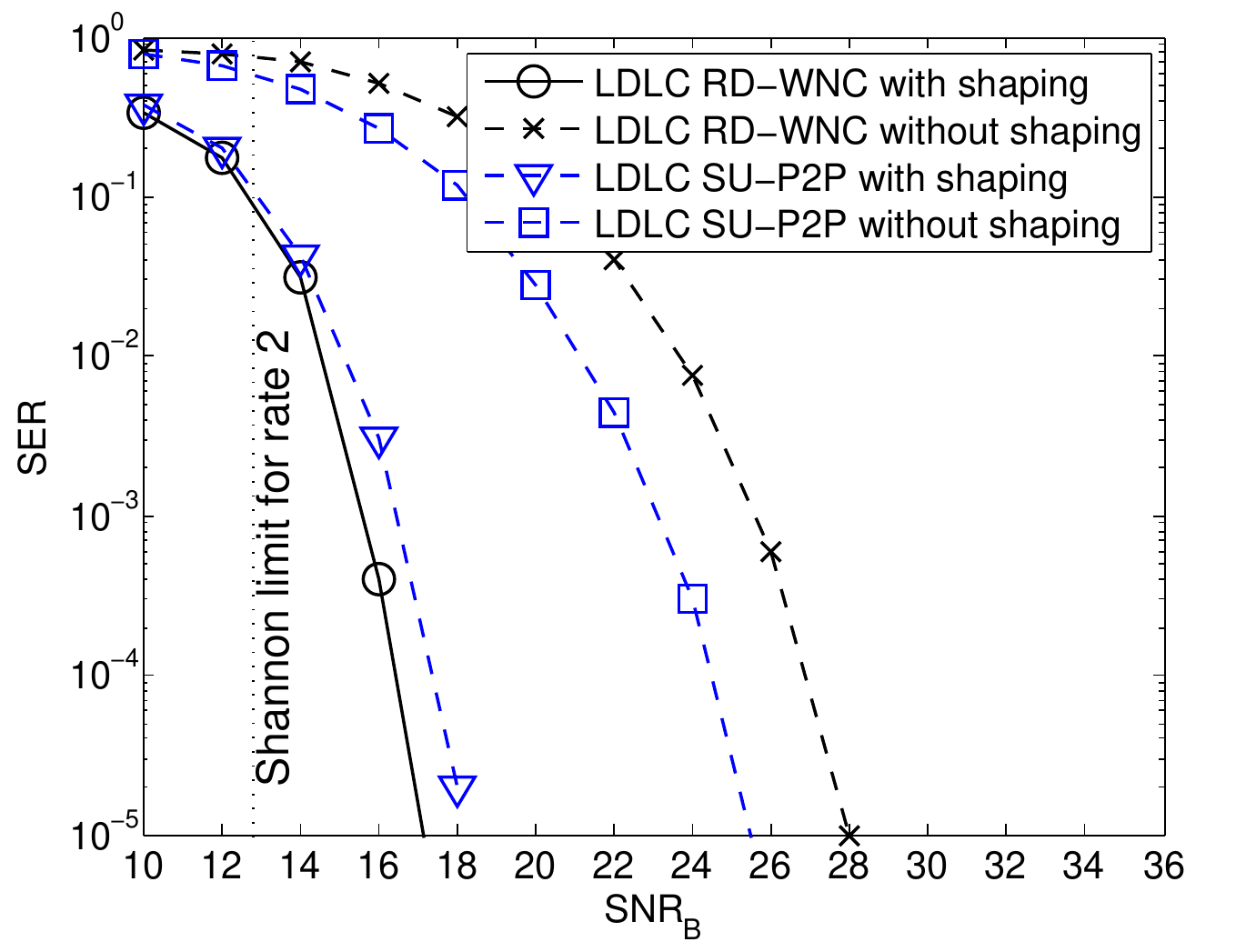}
	\caption{The SER results of LDLC RD-WCN and LDLC SU-P2P for user B.} \label{ldlc_ub}
	\vskip -0.1in
\end{figure}

We obtain the symbol error rates (SER) of user A, $p\left( {{{\widehat b}_{A,i}} \ne {b_{A,i}}} \right)$, and user B, $p\left( {{{\widehat b}_{B,i}} \ne {b_{B,i}}} \right)$, by averaging the decoding results over 1000 codewords for each user. The SER results of user A and user B in terms of SNR are shown in Fig.~\ref{ldlc_ua} and Fig.~\ref{ldlc_ub}, respectively. From the simulation results, we have the following observations:

First, the simulation results indicate that the LDLC implementation of our framework for rate-diverse wireless network coding (RD-WNC) can indeed achieve the same performance as the LDLC for single-user point-to-point (SU-P2P) channels. Thus, as the codeword length goes to infinite and the lattice generate by LDLC approaches the theoretically optimal lattice as discussed in Section III, the performances of LDLC SU-P2P  will approach the Shannon limit \cite{LDLC2008}, as well as that of LDLC RD-WNC.

Second, consistent with our statement in Section IV, the shaping gain of LDLC RD-WNC (around 10 dB) is larger than that of LDLC SU-P2P (5 dB). 

Third, for user B in Fig. 11 (i.e., the user with the lower rate), LDLC RD-WNC has an additional  $10{\log _{10}}\left( {{6 \mathord{\left/
			{\vphantom {6 5}} \right.
			\kern-\nulldelimiterspace} 5}} \right) \approx 0.8$ dB shaping gain compared with LDLC SU-P2P.  This is also consistent with our analysis in Section IV.C.

In conclusion, using implementable practical LDLC codes, we have empirically validated the feasibility and optimality of our framework for RD-WNC, which was shown to be theoretically optimal in Section III.

\subsection{The simulations of BICM Rate-Diverse Wireless Network Coding}

This subsection presents simulation results on the BER performance of our BICM framework for rate-diverse wireless network coding (BICM RD-WCN) in Section V.  In all simulations, the channel code used is the regular Repeat Accumulate (RA) code \cite{abbasfar2004accumulate}.  We can simply change the repetitions of the RA code to obtain different code rates. The decoding algorithm for RA codes is the iterative BP algorithm. The RA decoder performs 20 BP iterations to ensure convergence. The data packets for users A and B contain $10^4$  and  $5\times10^3$ binary source bits, respectively. Thus, the source rate for user A is twice that for user B. We compare our BICM framework with other schemes proposed for rate-diverse wireless network coding \cite{chen2010novel, yun2010rate}.

We first compare BICM RD-WCN with the joint modulation scheme in \cite{chen2010novel}. The joint modulation scheme originally proposed in \cite{chen2010novel} did not consider the effect of channel coding. In \cite{chen2010novel}, the modulator of BS maps two bits targeted for user A and one bit targeted for user B (totally three bits) into a point of an 8-PSK constellation. The 8-PSK constellation labeling map is optimized to have large inter-point Euclidean distances at the receivers \cite{chen2010novel}.  The joint 8-PSK constellation and its labeling map proposed by \cite{chen2010novel} can be found in Fig.1 (c) of \cite{chen2010novel}. At the receiver of user A, using the 1-bit side information, the 8-PSK constellation is reduced to an equivalent QPSK constellation. At the receiver of user B, using the 2-bit  side information, the 8-PSK constellation is reduced to an equivalent BPSK constellation. Although \cite{chen2010novel} focused on modulation and did not consider channel coding, it is straightforward to incorporate channel codes into the system. In our investigation here, for the joint modulation scheme, we apply an RA code with code rate $r = {1 \mathord{\left/
		{\vphantom {1 2}} \right.
		\kern-\nulldelimiterspace} 2}$
to the binary source bits of both user A and B. The coded bits are interleaved, and then the joint 8-PSK modulation is performed according to the constellation labeling map shown in Fig.1 (c) of \cite{chen2010novel}.    

\begin{figure}[!t]
	\centering
	\includegraphics[width=3in]{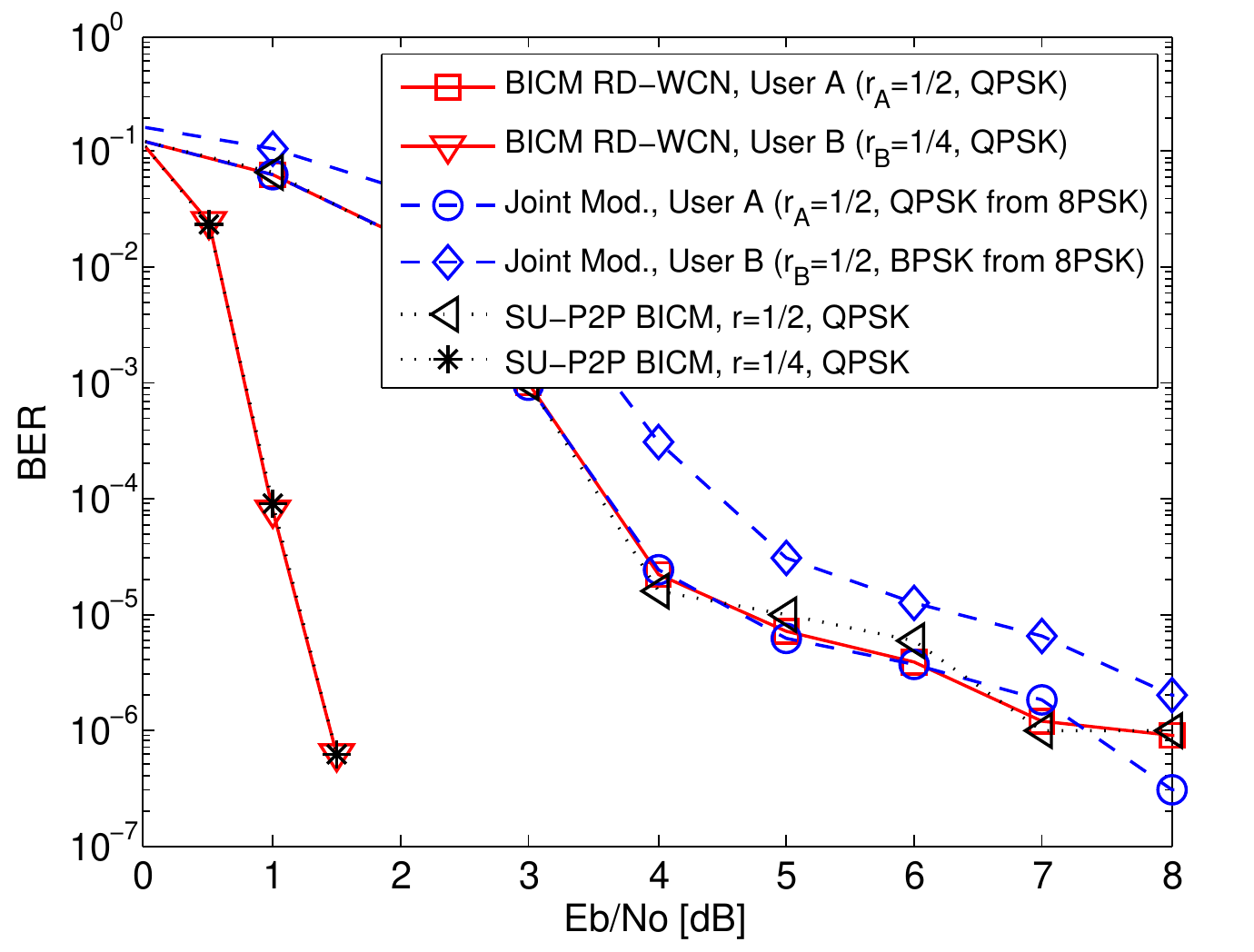}
	\caption{The BER results of our BICM RD-WCN and the joint modulation scheme \cite{chen2010novel}. } \label{jointmod}
	\vskip -0.1in
\end{figure}

Unlike the joint modulation scheme, our BICM RD-WCN applies channel codes with different code rates to the two users, as illustrated in Fig.~\ref{bicm_system_model}. Specifically, RA codes with code rates ${r_A} = {1 \mathord{\left/
		{\vphantom {1 2}} \right.
		\kern-\nulldelimiterspace} 2}$
and ${r_B} = {1 \mathord{\left/
		{\vphantom {1 4}} \right.
		\kern-\nulldelimiterspace} 4}$ are used to encode the binary source bits of user A and user B, respectively. This results in the same total number of coded bits for users A and B. Bit-wise XOR are then applied to the coded bits of users A and B. The XORed bits are then interleaved, and the standard QPSK modulation with gray labeling map is applied on the interleaved XORed bits. Therefore, both BICM RD-WCN and the joint modulation scheme serve user A at the data rate of 1 bit per channel use and user B at the data rate of half bit per channel use. The BER results of BICM RD-WCN and the joint modulation scheme are presented in Fig.~\ref{jointmod}. As the benchmarks, the performances of the SU-P2P BICM schemes (the $r = {1 \mathord{\left/
		{\vphantom {1 2}} \right.
		\kern-\nulldelimiterspace} 2},{1 \mathord{\left/
		{\vphantom {1 4}} \right.
		\kern-\nulldelimiterspace} 4}$
rate RA codes combined with the standard QPSK with gray labeling map) are also given in Fig.~\ref{jointmod}.  

In Fig.~\ref{jointmod}, we can see that indeed BICM RD-WCN has the same performance as SU-P2P BICM; the joint modulation scheme and BICM RD-WCN have the same performance for user A; BICM RD-WCN gives better performance for user B than the joint modulation scheme does (around 3.5 dB gain at BER of ${10^{ - 4}}$).

For the joint 8-PSK modulation scheme, the equivalent QPSK at the receiver of user A is a standard QPSK constellation with gray labeling map. Therefore, it is easy to see that, for user A, the joint modulation scheme can achieve the same BER performance as BICM RD-WCN and SU-P2P BICM under the setup of the rate-${1 \mathord{\left/
		{\vphantom {1 2}} \right.
		\kern-\nulldelimiterspace} 2}$ RA code and the standard QPSK constellation.

The 3.5 dB gain for user B obtained by BICM RD-WCN over the joint modulation scheme comes from two parts. The first part is the 0.7 dB gain from the modulation; the second part is the 2.8 dB gain from the channel coding. For the joint 8-PSK modulation, the constellation of the equivalent BPSK at the receiver of user B is not a standard BPSK constellation. It has an inter-point Euclidean distance ${d_2} = 1.85$ (see in Fig.~1 (c) of \cite{chen2010novel}). However, the inter-point Euclidean distance of the standard BPSK constellation is ${d_{BPSK}} = 2$. This implies that compared with the standard BPSK constellation, there is a $10{\log _{10}}\left( {{{d_{BPSK}^2} \mathord{\left/
			{\vphantom {{d_{BPSK}^2} {d_2^2}}} \right.
			\kern-\nulldelimiterspace} {d_2^2}}} \right) \approx 0.7$
dB performance loss in the equivalent BPSK constellation at the receiver of user B. It is well-known that the BER performances of standard BPSK and QPSK are the same in terms of ${E_b}/{N_0}$ \cite{proakisdigital}. And, the modulation of our BICM framework is the standard QPSK. Thus, the first 0.7 dB gain in user B's BER performance by BICM RD-WCN over the joint modulation scheme is from the modulation. Indeed, we can observe this 0.7 dB gain in Fig.~\ref{jointmod} (the gain of the SU-P2P BICM scheme with $1/2$ rate RA code and QPSK modulation over the joint modulation scheme for user B). Our BICM RD-WCN accommodates the different data rates by adding more redundancies to the channel-cooed bits of the user with lower rate. In our setup, the channel code of user B is the rate-$1/4$ RA code for BICM RD-WCN and the rate-$1/2$ rate RA code for the joint modulation scheme. From the performances of SU-P2P BICM in Fig.13, we know that the $1/4$ rate RA code can provide about 2.8 dB more coding gain than $1/2$ rate RA code does at BER of ${10^{ - 4}}$. Therefore, compared with the joint modulation scheme, BICM RD-WCN obtains this second 2.8 dB gain from the channel coding.

\begin{figure}[!t]
	\centering
	\includegraphics[width=3in]{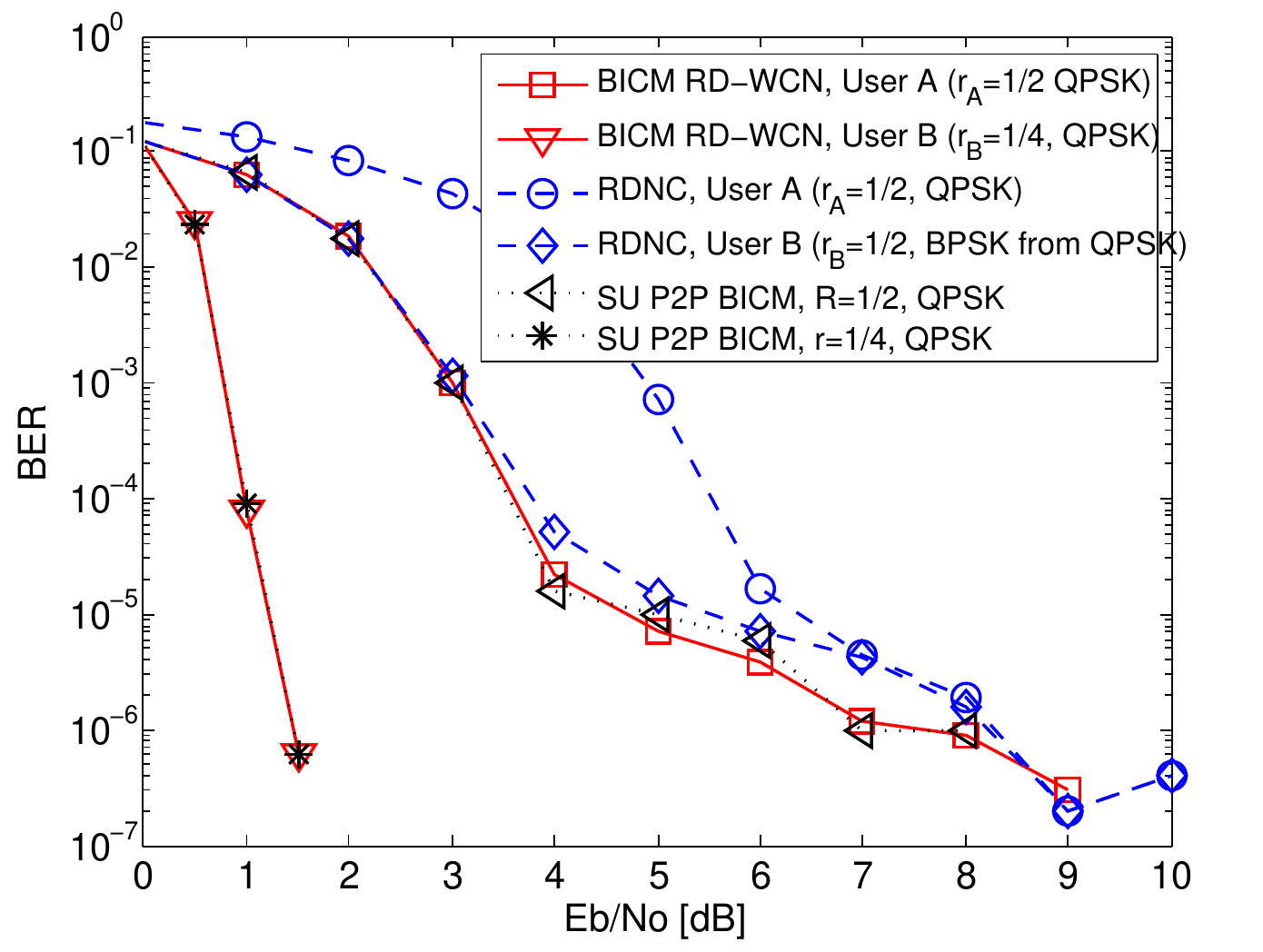}
	\caption{The BER results of our BICM RD-WCN and RDNC with QPSK in \cite{yun2010rate}.} \label{qpsk}
	\vskip -0.1in
\end{figure}

\begin{figure}[!t]
	\centering
	\includegraphics[width=3in]{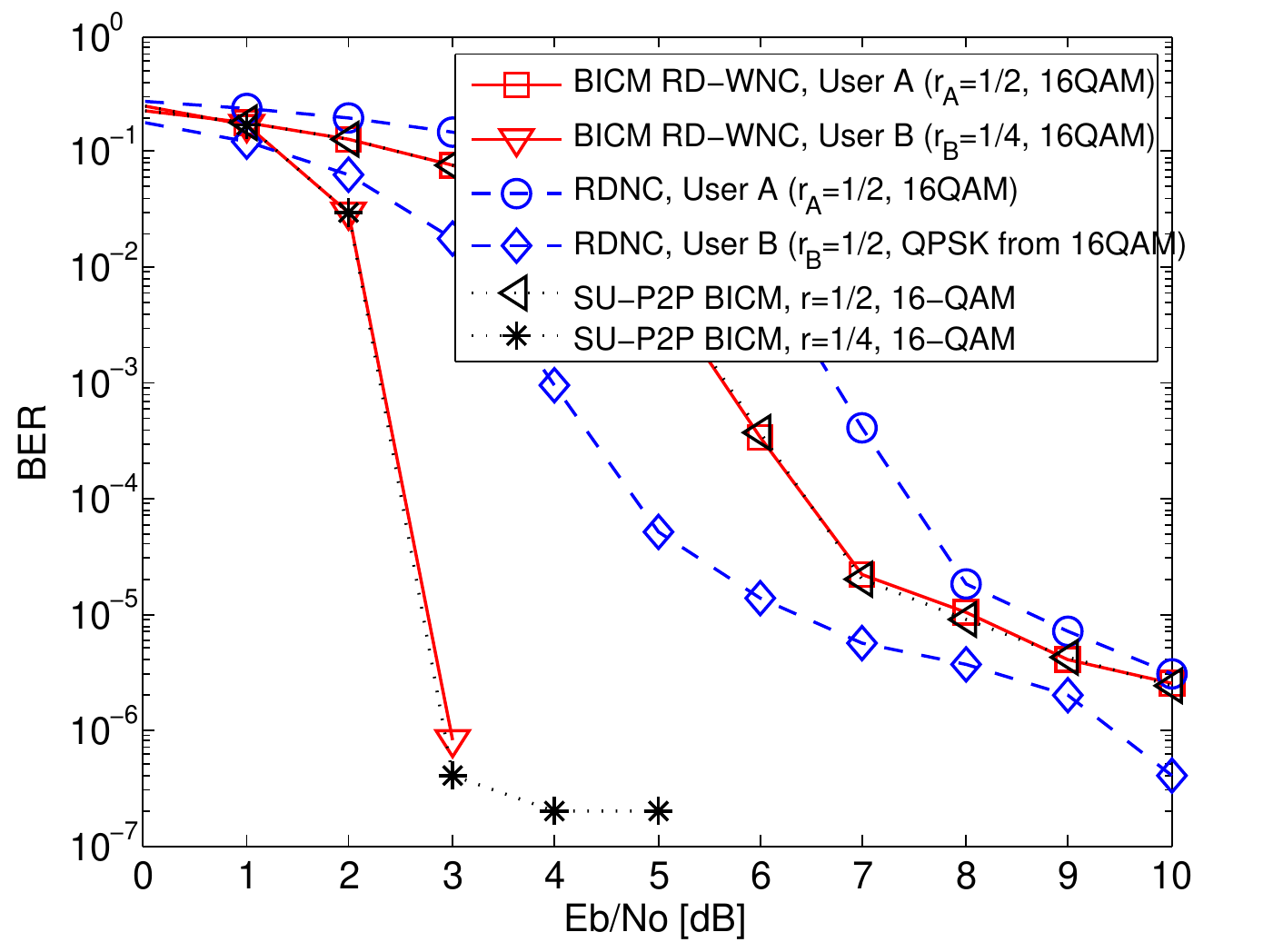}
	\caption{ The BER results of our BICM RD-WCN and RDNC with 16QAM in \cite{yun2010rate}.} \label{16qam}
	\vskip -0.1in
\end{figure}

We next compare our BICM RD-WCN with the rate-diverse network coding (RDNC) scheme proposed in \cite{yun2010rate}. Assuming the two users employ channel codes of the same rate, RDNC in \cite{yun2010rate} just deals with rate diversity within the modulation process. We briefly review RDNC below. The binary source bits of user A and B are encoded into channel-coded bits using a channel code of rate  $r$. The vectors of coded bits for user A and B are   ${{\bf{c}}_A}$ and ${{\bf{c}}_B}$. The length of ${{\bf{c}}_A}$  is the double of that of ${{\bf{c}}_B}$. Then, the coded symbols are interleaved to obtain the interleaved bits: ${\widetilde{\bf{c}}_A} = \Pi \left( {{{\bf{c}}_A}} \right)$, ${\widetilde{\bf{c}}_B} = \Pi \left( {{{\bf{c}}_B}} \right)$. Then, to make the vectors of the interleaved bits have equal lengths, RDNC inserts zeros within ${\widetilde{\bf{c}}_B}$ in an alternate manner: ${\widetilde c_{B,1}},0,{\widetilde c_{B,2}},0,{\widetilde c_{B,3}},0, \cdots $, where $0$s are the inserted zero bits.  The network coding of RDNC is the bit-wise XOR of ${\widetilde{\bf{c}}_A}$ and  the zero-inserted ${\widetilde{\bf{c}}_B}$. Finally, the network-coded bits are mapped to the constellation points of the modulation according to a special designed labeling map \cite{yun2010rate}.  

We investigate the BER performances of QPSK RDNC and QPSK BICM RD-WCN. The QPSK constellation labeling map for RDNC was developed in \cite{yun2010rate} (see Fig. 4 of \cite{yun2010rate}).  Every two network-coded bits are mapped to one QPSK constellation point according to the specially designed constellation labeling map in \cite{yun2010rate}. For QPSK RDNC,  a rate-$1/2$ RA code is applied to the source messages of  users A and B. For QPSK BICM RD-WCN, a rate-$1/2$  RA code is applied to user A, and a rate-$1/4$ RA code is applied to user B. The BER results are given in Fig.~\ref{qpsk}. We can see that, for both of users A and B, BICM RD-WCN gives better BER performance than RDNC does. For RDNC, the receiver of user A obtains a QPSK constellation after subtracting its side information; however, the QPSK constellation labeling map at the receiver is not gray labeling anymore. BICM RD-WCN does not lose the optimality of the SU-P2P BICM framework and the gray labeling for QPSK remains valid at the receivers. Due to this superiority in modulation, BICM RD-WCN gives better BER than QPSK RDNC does for user A, For RDNC, the receiver of user B obtains a standard BPSK constellation after substracting its side information; however, the channel coding of user B operates at a higher rate compared to BICM RD-WCN. Due to this superiority in channel coding, BICM RD-WCN also gives better BER than QPSK RDNC does for user B.

We next consider the 16QAM modulation for BICM RD-WCN and RDNC. The 16QAM constellation labeling map for RDNC was developed in \cite{yun2010rate} (see Fig. 6 of \cite{yun2010rate}).  Every four network-coded bits are mapped to one 16QAM constellation point according to the special designed constellation labeling map in \cite{yun2010rate}. The 16QAM constellation labeling for BICM RD-WCN is still the gray labeling. For 16QAM RDNC, a rate-$1/2$  RA code is applied to users A and B. For 16QAM BICM RD-WCN, a rate-$1/2$ RA code is applied to user A, and a rate-$1/4$ RA code is applied to user B. The above setups of 16QAM RDNC and 16QAM BICM RD-WCN also give the same number of coded bits for both users. The BER results are given in Fig.~\ref{16qam}. As the benchmarks, the performances of SU-P2P BICM schemes (the $r = {1 \mathord{\left/
		{\vphantom {1 2}} \right.
		\kern-\nulldelimiterspace} 2},{1 \mathord{\left/
		{\vphantom {1 4}} \right.
		\kern-\nulldelimiterspace} 4}$
RA codes with rates  are combined with the gray labeling 16QAM) are also given in Fig.~\ref{16qam}. We can see that our 16QAM BICM RD-WCN still has the same performance as 16QAM SU-P2P BICM; for both of user A and B, 16QAM BICM RD-WCN gives better BER performances than 16QAM RDNC does. The reasons are the same as the reasons for the comparison between QPSK BICM RD-WCN and QPSK RDNC.  

From the above simulation studies, we can conclude that our BICM RD-WCN can accommodate the different source rates of the users without losing the optimality of SU-P2P BICM for both users; other schemes proposed for rate-diverse scenarios cannot guarantee the optimality for both users simultaneously.

\section{Conclusion}

We have proposed a nested-lattice-code encoding/decoding framework to achieve the optimal capacity pair of wireless broadcast channels with side information at the users. Although the nested-lattice-code framework is optimal theoretically, its exact implementation faces many difficulties. In particular, the lattice quantization decoding, which searches for the transmitted lattice point over the lattice space, has complexity that grows exponentially with codeword length N. Fortunately, with the principle of virtual single-user channels suggested by our framework, we can implement our framework using practical codes with implementable decoding algorithms. To illustrate this, we implemented our encoding/decoding framework using LDLC that can be decoded using a practical BP algorithm. We further applied the design principle to a BICM scheme, where binary channel codes are used. Simulation results of the LDLC and BICM implementations indicate that, for rate-diverse wireless network coding systems, our encoding/coding framework indeed can enable the two users to simultaneously achieve/approach their respective channel capacities.

\ifCLASSOPTIONcaptionsoff
  \newpage
\fi

\bibliographystyle{ieeetr}
\bibliography{database}

\end{document}